\documentclass[10pt]{article}

\usepackage{doublespace}
\usepackage{epsfig}
\usepackage{latexsym}
\usepackage{amsmath}

\def\mtitle#1{{\Large\bf\centering\ignorespaces#1\vskip2.5pt}}
\def\mauthor#1{{\centering\ignorespaces#1\\[0.4ex]}}
\def\maddress#1{{\it \centering\ignorespaces#1\\[0.4ex]}}
\def\mdate#1{{\centering\ignorespaces#1\vskip2.5pt}}

\def\inout{{\rm in} \ra {\rm out}}

\def\>{\rangle}
\def\<{\langle}

\def\tr{{\rm tr}}
\def\non{\nonumber}

\def\lbL{ \left[\rule{0pt}{2.4ex}\right. }
\def\rbL{ \left.\rule{0pt}{2.4ex}\right] }
\def\lbm{ \left[\rule{0pt}{2.1ex}\right. }
\def\rbm{ \left.\rule{0pt}{2.1ex}\right] }

\def\lpL{ \left(\rule{0pt}{2.4ex}\right.\! }
\def\rpL{ \!\left.\rule{0pt}{2.4ex}\right) }
\def\lpm{ \left(\rule{0pt}{2.1ex}\right. }
\def\rpm{ \left.\rule{0pt}{2.1ex}\right) }

\def\mchi{\raisebox{0.3ex}{$\chi$}}
\def\tchi{\raisebox{0.3ex}{$\chi$}~}
\def\ss{\hspace*{0.15ex}}

\def\ot{\otimes}
\def\la{\leftarrow}
\def\ra{\rightarrow}
\def\lra{\leftrightarrow}

\renewcommand{\sec}[1]{Sec.~\ref{sec:#1}}
\newcommand{\eq}[1]{Eq.~(\ref{eq:#1})}

\def\be{\begin{equation}}
\def\ee{\end{equation}}
\def\bea{\begin{eqnarray}}
\def\eea{\end{eqnarray}}
\def\ben{\begin{eqnarray*}}
\def\een{\end{eqnarray*}}

\newtheorem{theorem}{Theorem}

\newtheorem{definition}[theorem]{Definition}

\newcommand{\qed}{\hspace*{1ex}{\large $\Box$}}

\begin{document} 

\mtitle{On the capacities of bipartite Hamiltonians and unitary gates}
\mauthor{
C.~H.~Bennett,$^1$ 
A.~W.~Harrow,$^{2}$\footnote[3]{Correspondence:
aram@mit.edu,~wcleung@cs.caltech.edu} 
D.~W.~Leung,$^{13}$ 
J.~A.~Smolin$^1$
}
\maddress{
$^1$IBM TJ Watson Research Center, P.O. Box 218, Yorktown Heights, NY
10598, USA \\
$^2$MIT Physics Dept., 77 Massachusetts Avenue, Cambridge, MA 02139, USA
}
\mdate{\today}

\abstract{We consider interactions as bidirectional channels.  We
investigate the capacities for interaction Hamiltonians and nonlocal
unitary gates to generate entanglement and transmit classical
information.  We give analytic expressions for the entanglement
generating capacity and entanglement-assisted one-way classical
communication capacity of interactions, and show that these quantities
are additive, so that the asymptotic capacities equal the
corresponding 1-shot capacities.  We give general bounds on other
capacities, discuss some examples, and conclude with some open
questions.}

\setlength{\parindent}{0ex}
\setlength{\parskip}{1.5ex}

\section{Introduction}
\label{sec:intro}

\subsection{Motivation I -- converting given interactions into abstract 
resources}
\label{sec:introgen}

The fundamental physical resource for performing various quantum
information processing tasks is the interaction between various
quantum systems.
These quantum systems can be, for example, individual registers in a
quantum computer, or systems possessed by isolated parties. 
%
The interactions are Hamiltonians or their discrete time incarnation
as unitary gates that are nonlocal with respect to individual systems.
The information processing tasks of interest include converting a {\em
given} interaction into a universal quantum gate, generating
entanglement between remote parties, and communicating classical or
quantum information.
%
%
These tasks output what are considered abstract resources in quantum
information theory, such as entanglement and classical communication. 
The study of the conversions among these various resources and the
efficiencies thereof has proved to be a fruitful field.

While our knowledge about the optimal use of a nonlocal interaction to
provide the derived resources is far from complete, important progress
has been made.
References \cite{Prehistory1,Prehistory2,Eisert00,Collins00} provide
motivating examples on the interconversion of these resources.  
Interconversion tasks can be classified as follows: 

(1) Simulation of one nonlocal Hamiltonian or gate by another: work
described in Ref.~\cite{simall} presents methods for doing so, some of
which are optimal under certain circumstances.

(2) Generation of entanglement using nonlocal Hamiltonians and gates:
Partial results are obtained in
Refs.~\cite{Zanardi00,Cirac00,Dur00,Kraus00}.

(3) Classical (or quantum) communications using nonlocal
Hamiltonians and gates.

(4) Performing a nonlocal quantum operation using entanglement and
classical communication: This is the converse of the last two tasks.
General formalisms for the $1$-shot bipartite case and 
methods for more specific gates are given in Ref.~\cite{Cirac00}. 
%

These tasks are related.  First, entanglement, forward classical
communication, and backward classical communication are strictly
incomparable resources: no one of them can be generated even from an
infinite supply of the other two.
Thus the capacity of a given interaction to create each of the three
resources cannot exceed the amount used to perform the interaction.
For example, the {\sc cnot} can be simulated using $1$
ebit\footnote{The unit ebit is defined to be the amount of
entanglement in the EPR state ${1 \over \sqrt{2}} (|00\>+|11\>)$.},
$1$ forward, and $1$ backward classical bit~\cite{Gottesman98}, so
that the entanglement capacity and both forward and backward classical
capacities are upper bounded by $1$.
Second, the efficiency for one interaction to simulate another
provides bounds on the relative efficiency for the interactions to 
generate resources.   
For example, any capacity of {\sc swap} is at most $3$ times that of
{\sc cnot} since the {\sc swap} can be written as $3$ {\sc cnot}s.  

In this paper, we focus on tasks (2) and (3), and investigate the
capacities of a unitary interaction to generate entanglement and
perform classical communication.
The unitary interaction can be a nonlocal Hamiltonian or gate.  
We are primarily concerned with the asymptotic limit, when many 
uses of the gate (or a long duration of the Hamiltonian) are given.
We consider an interaction on two $d$-dimensional systems, allow
unlimited local operations, local ancillas of arbitrarily large
dimensions, and arbitrary input state.
We give expressions for the entanglement generating capacity and the
entanglement-assisted one-way classical
capacity~\cite{Holevo98,Schumacher97,Shor01,Holevo01} of an
interaction.
We show that these quantities are additive in the sense that the
amount of entanglement or classical communication generated by $n$
uses of a gate is $n$ times the amount generated by one use.

\subsection{Motivation II -- interactions as bidirectional channels} 
\label{sec:bichan}

The capacities of generating entanglement and communication are well
studied in the context of a {\em noiseless or noisy} channel
connecting a sender (Alice) to a receiver (Bob).
In this usual model of a quantum channel, a quantum system is
physically transported from Alice to Bob, with possible changes
(noise) caused by a quantum operation~\cite{Nielsen00} (i.e. a
trace-preserving, completely positive, or TCP, linear map).
%
%
This model of a channel is unidirectional -- Bob cannot send
information to Alice.
%
%
However, such unidirectional interactions are a special case of
quantum interactions, and in general, a quantum system cannot affect
another without being changed itself.
For example, the {\sc cnot} (defined in the computational basis)
operates in reverse direction in the conjugate basis, and transmits an
equivalent amount of information in either direction when used in
conjunction with other local gates.  
%

In view of this, we generalize the usual model of a quantum channel to
take into account the bidirectional nature of a quantum interaction.
We define a ``bidirectional channel'' as a bipartite quantum
operation.  Alice and Bob each inputs a state to the ``bidirectional
channel'' and receives an output.
This work can be viewed as studying the entanglement and classical
capacities in bidirectional channels, restricted to the unitary case.

Throughout this paper a {\em protocol} means a procedure that
uses a nonlocal gate one or more times, or a nonlocal Hamiltonian
for some total amount of time,  possibly also consuming and/or
producing various amounts of other standard resources, such as
entanglement and classical communication in each direction.
We always allow unlimited local operations, and we are interested
in a protocol's {\em net} yield (production minus consumption) of
standard resources per gate use or unit interaction time.
The protocol can be written as a quantum circuit, and the net effect
can be described as a bipartite quantum operation, with bipartite
input and output.  We call this quantum operation the protocol as
well.
In general there is a tradeoff among the yields of various resources 
when the protocol is varied.  
For example, {\sc cnot} can transmit a classical bit in the forward or
backward direction, but not both.
As back communication is generic in a bidirectional channel, a
protocol using it is generically interactive.

In the next two subsections, we provide more detailed introductions to
the two tasks studied in this paper and discuss closely related work.

\subsection{Entanglement generating capacity of bidirectional channels}
\label{sec:introecap}

In Ref.~\cite{Bennett96a}, the quantum communication capacity of a channel
is shown to be equal to its capacity for generating pure entanglement;
a greater quantum capacity typically results if two-way classical
communication is allowed.
Likewise, a bidirectional channel (bipartite quantum operation) can be 
used to generate entanglement. 
Simple examples are considered in
Refs.~\cite{Prehistory1,Prehistory2,Eisert00,Collins00}.
Reference~\cite{Zanardi00} considers the average amount of entanglement
created by one use of a nonlocal operation on a distribution of
product states.
Reference~\cite{Cirac00} classifies the type of entanglement (bound or
distillable) that can be created from product states.
Reference~\cite{Dur00} considers the optimal $1$-shot rate of creating
entanglement using an arbitrary $2$-qubit Hamiltonian on possibly
entangled pure input states without local ancillas.
Reference~\cite{Kraus00} considers the optimal amount of entanglement
created by one use of an arbitrary $2$-qubit gate on pure product
input states without ancillas.
References \cite{Dur00,Kraus00} also exhibit examples in which
local ancillas increase the amount of entanglement created.

In this paper, we follow the philosophy of Ref.~\cite{Bennett96a} and
investigate the asymptotic entanglement generating capacity of a
bidirectional channel acting on two $d$-dimensional systems.
Contrary to previous work~\cite{Zanardi00,Cirac00,Dur00,Kraus00}, we
do not restrict ourselves to qubit systems, we allow arbitrary local 
ancillas and input states (including entangled or mixed states), and
we consider the most general asymptotic protocols.
We also consider the effect of many auxiliary resources including 
classical communication.  
We restrict our attention to unitary bidirectional channels.  We
derive the expression for the capacity, show that it is additive, and
discuss the optimal protocol.

%
Leifer, Henderson, and Linden~\cite{Leifer02} have independently
shown, by similar arguments, that the asymptotic entanglement
generating capacity on pure input states is an optimization over a
$1$-shot expression.  They also investigate the capacities for many
$2$-qubit gates with low dimension ancillas both analytically and
numerically.

\subsection{Classical communication capacities of bidirectional channels}
\label{sec:introcgen}

The classical capacity of an ordinary (unidirectional) quantum channel
is in general affected by the availability of auxiliary resources,
such as entanglement~\cite{Bennett92} and back communication.
For a general noisy quantum channel, the capacity without auxiliary
resources is found in Refs.~\cite{Holevo98,Schumacher97}, and that
with unlimited supply of pure entanglement is found in 
Refs.~\cite{Shor01,Holevo01}.
The capacity for a noiseless quantum channel with unlimited supply of
a certain noisy entangled state is found
in Refs.~\cite{Bose00,Hiroshima00,Bowen01,Horodecki01sd}.

In treating bidirectional channels, we again follow the philosophy of
Refs.~\cite{Holevo98,Schumacher97,Shor01,Holevo01} and consider
various asymptotic classical capacities of unitary bidirectional
channels of arbitrary dimensions.  We allow unlimited local resources,
including free instantaneous local operations, and the freedom for
Alice and Bob to attach and remove local ancillas.  Shared randomness
is also given as a resource.  Our philosophy is also similar to
Shannon's study of the classical capacities of classical two-way
communication channels~\cite{Shannon61}.

A new ingredient in the case of bidirectional channels is the
simultaneous forward and backward communication, resulting in a {\em
pair} of achievable rates.  One can define many classical capacities
other than the forward and the backward capacities.  Generally, there
is a tradeoff between the forward and backward rates.

Our long term goals are to obtain expressions for these capacities,
understand the tradeoff between forward and backward communication,
and relate the quantities to other capacities such as the entanglement
generating capacity.
In this paper, we define various asymptotic capacities of bidirectional
channels.
We obtain an expression for the one-way (forward or backward)
entanglement-assisted classical capacity for any arbitrary nonlocal
gate or Hamiltonian, and the protocol achieving it.
The asymptotic capacity is achieved by a $1$-shot expression, as an
optimization over input ensembles for one use of the gate.  

%
We remark that other independent investigations on optimal methods to
perform classical communication in low dimensions without entanglement
assistance are being conducted~\cite{Hammerer02,Bristol,Bernstein02}.

\subsection{Structure and assumptions of the paper} 
\label{sec:structure}

In the next section, we discuss in detail the problem of entanglement
generation, and derive the expression for the entanglement generating
capacity for any nonlocal gate.  In Sec.~\ref{sec:def}, we define
various classical communication capacities, followed by a derivation
of the entanglement-assisted one-way classical capacity for any nonlocal
gate in Sec.~\ref{sec:ce}.  
We discuss the similarities and differences between the two derivations 
in \sec{diss}.  
In Sec.~\ref{sec:gbdd}, we prove various general bounds relating the
capacities for entanglement generation and classical communication. 
We conclude in Sec.~\ref{sec:evilgates} with open questions and
examples of unitary bidirectional channels.

Throughout the paper, we assume the following.  
Unless otherwise noted, logarithms are in base 2.  $O(n)$ and
$o(n)$ respectively denote functions linear and sublinear in $n$.
$U$ denotes a nonlocal gate acting on two $d$-dimensional systems
(with shorthand $d \times d$) in the possession of Alice and Bob.
%
%
They have access to the following local resources: 
\begin{quote}
Local ancillas of arbitrarily large but finite dimensions and
unlimited local operations. 
\end{quote}
We do not consider ancillas of infinite dimensions and do not know if
they can be more useful.

Though we have motivated the discussion with both Hamiltonians and
gates, we now argue that it is sufficient to focus on gates only.
This is because Hamiltonian capacities are simply gate capacities in
the limit of infinitesimal gates, so that any Hamiltonian capacity 
can be obtained from the corresponding gate capacity. 
A protocol using a Hamiltonian is similar to one using a gate, with
additional freedom on how long each free Hamiltonian evolution can
last before being interspersed with local operations.
However, different durations of evolution are simply concatenation of
different numbers of infinitesimal ones.  
Thus any Hamiltonian capacity $G_H$ can be expressed in terms of the
corresponding gate capacity $G_U$:
$$G_H = \lim_{s \ra 0} \; {1 \over s} \; G_{U = e^{-iHs}} \,.$$ 


\section{Entanglement capacity of bidirectional channels}
\label{sec:ecap}

\subsection{Main idea} 

Before a formal treatment of the entanglement capacity, we first
illustrate our central idea with the following example.
Let $E_e$ be the entropy of entanglement~\cite{Bennett95}.  
Suppose the goal is to {\em increase} $E_e$ as much as possible.
Different uses of $U$ can be used sequentially or in parallel, and be
interspersed by LOCC (Local Operation and Classical Communication).
We allow an arbitrary pure input state with ancillas, possibly
entangled over different uses of $U$.
What is the optimal strategy?  
The answer turns out to be very simple.  
Consider the quantity 
\bea
	\Delta E_U \equiv \sup_{|\psi\>_{ABA'\!B'}} E_e(U_{AB}
	|\psi\>_{ABA'\!B'}) - E_e(|\psi\>_{ABA'\!B'})
\,, 
\label{eq:eentropy}
\eea
which represents the entanglement generated by optimizing the input
state for one use of $U$.
Let $|\psi_*\>_{ABA'\!B'}$ attain the supremum.  
Then applying individual uses of $U$ to copies of $|\psi_*\>_{ABA'\!B'}$ 
is asymptotically optimal.
This is because the total increase in $E_e$ in any asymptotic protocol
is at most the sum of the increases due to each use of $U$, each is no
greater than $\Delta E_U$.

In the following, we will develop this idea rigorously in the most
general setting.  We consider mixed input states and different
entanglement measures, and analyze the roles of various auxiliary
resources.

\subsection{Definitions and summary of results}

\label{sec:edefs}

The entanglement capacity of a gate $U$ can only be defined when the
entanglement measures for the input and output and the available
auxiliary resources are specified.

%
Traditionally, entanglement is a qualitative phenomenon.  The theory
of quantifying entanglement is not complete, though much progress has
been made~\cite{emeasall} (Refs.~\cite{Donald01,Vidal98} give
informative reviews).
Based on the transformation properties of entangled states, measures
of entanglement are defined which are very different in the asymptotic
and nonasymptotic regimes.  Different measures in the same regime
can also be inequivalent.
 
The entanglement generated by a protocol $\cal P$ on an input $\rho$ 
is intuitively 
\bea
	E_{\rm out}({\cal P}(\rho)) - E_{\rm in}(\rho)
\eea
where $E_{\rm in}$, $E_{\rm out}$ are the input and output entanglement
measures specified in the problem.  We can now define the entanglement 
capacity of $U$: 
\begin{quote}
\begin{definition} 
{\rm The $t$-shot entanglement capacity of $U$ is the maximum amount
of entanglement generated per use of of $U$ by any protocol ${\cal
P}_t$ that uses $U$ $t$ times, auxiliary resources labelled by $r$,
and local resources specified in Sec.~\ref{sec:structure} (ancillas 
of arbitrary but finite dimensions and unlimited local operations).
We consider two possible $t$-shot capacities, depending on the allowed
input state: \\
1. when the input is restricted to be a product state (without loss of
generality $|00\>$, since Alice and Bob can locally transform any
product state into any other product state):
\bea
	E_{\inout,U}^{(t,\emptyset,r)} \equiv 
	\sup_{{\cal P}_t} \; {1 \over t}  
	\lbm E_{\rm out}({\cal P}_t(|00\>\<00|)) - E_{\rm in}(|00\>\<00|) \rbm 
\label{eq:edeft0}
\eea
~~~~where the superscript $\emptyset$ denotes ``starting from
nothing'', and \\
2. when there is no restriction on the input state:  
\bea
	E_{\inout,U}^{(t,*,r)} \equiv 
	\sup_{{\cal P}_t} \; \sup_{\rho} \; {1 \over t}  
	\lbm E_{\rm out}({\cal P}_t(\rho)) - E_{\rm in}(\rho) \rbm 
\label{eq:edeft+}
\eea
~~~~where the superscript $*$ denotes an optimization over all possible 
input states.

The corresponding asymptotic capacities are:  
\bea
	E_{\inout,U}^{(\emptyset,r)} \equiv  
	\lim_{t \ra \infty} E_{\inout,U}^{(t,\emptyset,r)} ~~,~~~
	E_{\inout,U}^{(*,r)} \equiv  
	\lim_{t \ra \infty} E_{\inout,U}^{(t,*,r)} \,.
\label{eq:edef}
\eea
}
\end{definition}
\end{quote}

In our notation, we assume that an entanglement measure is written as
$E_x$ where the subscript $x$ labels the measure.  
For example, in $E_{\rm in}$, $E_{\rm out}$, and the notation for
the above capacities, the ``in'' and ``out'' are placeholders for the
entanglement measures being referred to.
Whenever $E_{\rm in} = E_{\rm out} = E_x$ we simplify the notation of
the capacity to $E_{x,U}^{(\cdots)}$.  Finally, an arbitrary
entanglement measure is written as $E$ without the subscript, and the
capacity is written as $E_{U}^{(\cdots)}$.

By definition, $E_{\inout,U}^{(t,\emptyset,r)} \leq
E_{\inout,U}^{(t,*,r)}$.
The capacity $E_{\inout,U}^{(t,*,r)}$ has an operational meaning that
a supply of the initial state $\rho$ is available at a price $E_{\rm
in}(\rho)$.  This is a resource, because the ability to create $\rho$
with an average cost $E_{\rm in}(\rho)$ is generally not guaranteed
(unless $E_{\rm in}$ is the entanglement cost~\cite{Hayden00}).
We refer to this as ``the resource $*$'' throughout the paper. 
In contrast, no such resource is assumed in the capacity
$E_{\inout,U}^{(t,\emptyset,r)}$.  

Since we are interested in asymptotic capacities, we are primarily
concerned with asymptotic measures.  These include the entanglement
cost $E_c$~\cite{Hayden00} and the distillable entanglement
$E_d$~\cite{Bennett96a}.  We also study the entanglement of formation
$E_f$~\cite{Bennett96a}, which is closely related to $E_c$.  All of
these measures coincide with the entropy of entanglement $E_e$ on pure
states.
%
%
As our results apply to more general measures, and may be useful in
other contexts, we follow an abstract
approach~\cite{Donald01,Vidal98}, which requires more technicalities
in our arguments.  However, the essence can be made clear by relating
to our simplified example, and we leave this step as an exercise to the
readers.

%
The auxiliary resources can be divided into three types according to
their quantities.
The first type is given in an amount that is negligible or can be
recovered at the end of the protocol.  For example, sublinear (in the
number of uses of $U$) amount of resources in the asymptotic case are
negligible, and catalytic resources in the $1$-shot case are used and
regenerated (for example, see Ref.~\cite{Jonathan99b}).
However, we need not consider these resources. 
In the $1$-shot case, catalytic resources are a subset of the resource
$*$ and need not be treated separately.
In the asymptotic case, sublinear amount of any resource can be
produced at a vanishing average cost and does not affect the
asymptotic capacity.
This is because any nonlocal gate has nonzero capacity to create pure
entanglement and to perform classical communication (see
Sec.~\ref{sec:gbdd}) from which any other resource can be produced.
The second type of resources are at least linear in the number of uses
of $U$.  To consider these resources is an important open question,
but it is out of scope of the present paper.
The third type of resources are unlimited and free.   
In the context of generating entanglement, we focus on the auxiliary
resource of unlimited $2$-way classical communication, labelled by
``{\rm cc}''.

%
Our results can be summarized in terms of the entanglement capacities
just defined.
\begin{itemize}
\item
In \sec{edd} we show that if $E_{\rm in} = E_{\rm out} = E$ and $E$
is nonincreasing under LOCC,
$E_{U}^{(1,*,{\rm cc})} \geq E_{U}^{(*,{\rm cc})}$.  
Thus given the resource $*$, the $1$-shot capacity is no less than the
asymptotic capacity.
We give a sufficient condition for additivity, $E_{U}^{(1,*,{\rm cc})}
= E_{U}^{(*,{\rm cc})}$, and we describe an optimal asymptotic protocol
that does not require classical communication (thus $E_{U}^{(t,*,{\rm
cc})} = E_{U}^{(t,*)}$).  
Additivity holds for many measures including $E = E_c, E_f, E_d$.
\item In \sec{asymres} we consider $E = E_c$, and we show that
$E_{c,U}^{(*,{\rm cc})} = E_{c,U}^{(\emptyset,{\rm cc})} =
E_{c,U}^{(\emptyset)}$ by describing an explicit protocol.  We show
the same for $E_f$.  In other words, $*$ or {\rm cc} does not
increase the asymptotic capacity of $E_{c,f}$, which can be attained
without classical communication nor a supply of the optimal input.
\item In \sec{diffmeas} we consider the maximum gain of pure
entanglement.  This is given by $E_{{\rm c} \ra {\rm d},U}^{(t,*,{\rm
cc})}$, and we show that it is equal to $E_{{\rm c},U}^{(t,*,{\rm
cc})}$.  Thus, the optimal protocol in \sec{asymres} applies, without
the need of resources $*$ or {\rm cc}.
\end{itemize}

\subsection{Expression for $E_{U}^{(*,\,{\rm cc})}$ when 
$E_{\rm in} = E_{\rm out} = E$}
\label{sec:edd}

Throughout this subsection, $E_{\rm in} = E_{\rm out} = E$ and both
resources $*$, {\rm cc} are available.

Let $|\Phi_n\> = {1 \over \sqrt{n}} \sum_{i=1}^n |i\>_A |i\>_B$ be 
the $n \times n$ maximally entangled state shared between Alice and Bob.

Unless otherwise stated, $E$ satisfies the following assumptions, but
is otherwise arbitrary: \\
A1. $E = 0$ for product states.  \\   
A2. $E$ is invariant under local unitaries. \\
A3. $E$ is nonincreasing under LOCC. \\
A4. $\forall_{\rho}$ $E(\rho \ot |\Phi_n\>\<\Phi_n|) = E(\rho) +
    E(|\Phi_n\>)$.  \\
A1-3 are basic axioms for entanglement measures, while A4 is needed to
define the ``net'' amount of entanglement generated by a protocol.  
Generally, we do not assume $E$ is normalized ($\forall_n$
$E(|\Phi_n\>) = \log n$) and will state the assumption explicitly when
it is needed.

We first state a lemma based on the following simple
observation~\cite{Eisert00,Collins00}.
Alice and Bob can implement $U$ if Alice teleports her input to Bob,
who applies $U$ locally in his own laboratory and teleports her output
to her.
This consumes two copies of $|\Phi_d\>$ and $2 \log d$ bits of classical
communication in each direction.

{\bf Lemma 1:} $\forall_t$ $E_U^{(t,*,{\rm cc})} \leq 2 E(|\Phi_d\>)$.  
Thus $E_U^{(*,{\rm cc})} \leq 2 E(|\Phi_d\>)$.   
If $E$ normalized, $E_U^{(*,{\rm cc})} \leq 2 \log d$.

{\bf Proof:} For any protocol with $t$ uses of $U$ and LOCC, modify it
by replacing each use of $U$ with its double teleportation
implementation.  Let $\rho_{\rm in}$ and $\rho_{\rm out}$ be the input
and output of the original protocol.  The modified protocol uses only
LOCC and has input $\rho_{\rm in} \ot |\Phi_d\>\<\Phi_d|^{\ot 2t}$ and
output $\rho_{\rm out}$.  Applying A3-4 to the modified protocol, $2 t
E(|\Phi_d\>) \leq E(\rho_{\rm out}) - E(\rho_{\rm in})$.  \qed

We now proceed to prove Theorem 1, which says that the asymptotic
capacity is equal to the $1$-shot capacity given the resources $*$ and
{\rm cc}.  This is done by proving two separate inequalities, each is
referred to as a half of the Theorem.

{\bf Theorem 1 (1st half):} 
$E_U^{(t,*,{\rm cc})} \leq E_U^{(1,*,{\rm cc})}$.

{\bf Proof:} Since LOCC operations cannot increase entanglement, the
best $1$-use protocol has the form:

\vspace*{-3ex}
\bea
\centering
\setlength{\unitlength}{0.60mm}
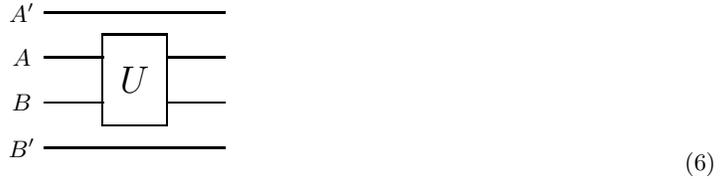
\begin{picture}(60,35)
%
\put(10,5){\line(1,0){40}}
\put(10,15){\line(1,0){13}}
\put(10,25){\line(1,0){13}}
\put(10,35){\line(1,0){40}}
\put(0,0){\makebox(10,10){$B'$}}
\put(0,10){\makebox(10,10){$B$}} 
\put(0,20){\makebox(10,10){$A$}} 
\put(0,30){\makebox(10,10){$A'$}} 
\put(37,15){\line(1,0){13}}
\put(37,25){\line(1,0){13}}
{\Large
\put(23,10){\framebox(14,20){$U$}}
}
\end{picture}
\label{eq:egen}
\eea
The only optimization is over the initial state, and thus 
\bea
	E_U^{(1,*,{\rm cc})} 
	= \sup_{\rho_{A\!A'\!B\!B'}} 
	  E( U_{A\!B} \; \rho_{A\!A'\!B\!B'} \; U_{A\!B}^\dagger ) 
	- E( \rho_{A\!A'\!B\!B'} ) \,.
\label{eq:e1}
\eea
In \eq{e1} and throughout the paper, the subscripts of an operator
denote the systems being acted on.
As an aside, $E_U^{(1,*,{\rm cc})} = E_U^{(1,*)}$ since no classical
communication is used in the protocol depicted in Fig.~(\ref{eq:egen}).

Now, consider any protocol with LOCC and $t$ uses of $U$.
Without loss of generality, we can divide the circuit into time steps
each having either $1$ use of $U$ or only LOCC operations.
The entanglement can only increase in the $t$ time steps with $U$ and
each is described by Fig.~(\ref{eq:egen}), by defining the ancillas
$A'$ and $B'$ to include all registers not acted on by $U$ in that
time step.
Thus, the total amount of entanglement generated by the protocol is no
more than $t E_U^{(1,*)}$, and $E_U^{(t,*,{\rm cc})} 
\leq E_U^{(1,*)} = E_U^{(1,*,{\rm cc})}$.\qed

We now consider sufficient conditions for additivity, 
$E_U^{(t,*,{\rm cc})} = E_U^{(1,*,{\rm cc})}$.
We say that $E$ is weakly additive on $\rho$ if $E(\rho^{\ot n}) = n
E(\rho)$.  $E$ is weakly additive if it is so on all $\rho$.  We say
that $E$ is strongly additive on $\rho_1$ if $\forall_{\rho_2}$
$E(\rho_1 \ot \rho_2) = E(\rho_1) + E(\rho_2)$.  $E$ is strongly
additive if $\forall_{\rho_1,\,\rho_2}$ $E(\rho_1 \ot \rho_2) =
E(\rho_1) + E(\rho_2)$.  
Weak and strong subadditivity and superadditivity are defined by
replacing the equality in the corresponding additivity definitions by
the inequalities $\leq$ and $\geq$ respectively.  

{\bf Theorem 1 (2nd half):} 
If $E$ is weakly additive or subadditive on the optimal {\em input} in
\eq{e1}, and is weakly additive or superadditive on the optimal {\em
output}, $E_U^{(t,*,{\rm cc})} \geq E_U^{(1,*,{\rm cc})}$.

{\bf Proof:} Consider the $t$-use protocol that repeats the $1$-use
protocol in Fig.~(\ref{eq:egen}) $t$ times, each on a separate copy of
the optimal input.  The entanglement generated is at least $t
E_U^{(1,*,{\rm cc})}$, and $E_U^{(t,*,{\rm cc})} \geq E_U^{(1,*,{\rm
cc})}$.\qed

Note that any measure which is weakly additive and is
nonincreasing under LOCC satisfies both halves of Theorem 1 and
$E_U^{(*,{\rm cc})} = E_U^{(1,*,{\rm cc})}$.
Weak additivity is not an axiomatic property of entanglement, and need
to be checked for individual measures before applying the 2nd half of
Theorem 1.  On the other hand, Theorem 1 does hold for most commonly
used measures.
Examples include $E_c$, $E_d$~\cite{edmono}, and the R\'{e}nyi entropy
(which includes the mixed state generalizations of the logarithm of the
Schmidt number and $E_e$ as special cases).  
It is an open problem whether $E_f$ is weakly additivity, however, we
will prove that the second half of Theorem 1 still holds for $E = E_f$.

In general, we say that ``Theorem 1 holds'' whenever both halves of
Theorem 1 hold.  \eq{e1} then provides an explicit expression for
$E_U^{(*,{\rm cc})}$ achieved by repeating the $1$-use protocol.  
In fact, the protocol only requires a supply of the optimal input (the
resource $*$) but not classical communication, and $E_U^{(*,{\rm cc})}
= E_U^{(*)}$.
In the following, we prove some lemmas for $E_U^{(1,*,{\rm cc})}$.  We
discuss how to obtain this supply of optimal input in \sec{asymres}.

In the expression for $E_U^{(1,*,{\rm cc})}$ in \eq{e1}, the supremum
is taken over finite but arbitrarily large dimensional ancillas
$A'B'$.  This can also be viewed as a limiting quantity as the ancilla
dimensions increase.

{\bf Lemma 2:} Suppose we restrict to $n$-dimensional $A',\,B'$ in
\eq{e1}, and denote the subsequent maximization by $e_n$.  Then, 
$\lim_{n \ra \infty} e_n = E_U^{(1,*,{\rm cc})}$.

{\bf Proof:} The sequence $\{e_n\}$ is increasing and bounded above by
Lemma 1.  Furthermore, $\forall_{\epsilon > 0}$ $\exists n_o$ such
that $e_{n_o} \geq E_U^{(1,*,{\rm cc})} - \epsilon$.  \qed

Lemma 2 provides one possible way to estimate the supremum in \eq{e1} 
suitable for numerical approaches.

We say that ``$E_U^{(1,*,{\rm cc})} - \epsilon$ can be attained on an
input $\rho$'' if the $1$-shot protocol generates an amount of
entanglement $E_U^{(1,*,{\rm cc})} \! - \epsilon$ on the input $\rho$.
The next $2$ lemmas show that the optimal input for the $1$-shot
capacity in \eq{e1} can be chosen pure for the specific measures $E_f$
and $E_c$.

{\bf Lemma 3:} 
For any $\epsilon > 0$, $E_{f,U}^{(1,*,{\rm cc})} - \epsilon$ can be
attained on a pure input state.  

{\bf Proof:} Let $\rho_{A\!A'\!B\!B'}$ be a state attaining the
supremum in \eq{e1} to within $\epsilon$.  We omit the system label when
it is $A\!A'\!B\!B'$ in Lemmas 3-4.
Let $\rho = \sum_i \lambda_i |\psi_i\>\<\psi_i|$ be an optimal
decomposition so that $E_f(\rho) = \sum_i \lambda_i E_f(|\psi_i\>)$.
Then,
\bea
	E_{fU}^{(1,*,{\rm cc})} - \epsilon & \leq & 
	E_f(U_{AB} \; \rho \; U_{AB}^\dagger) - E_f(\rho)
\non
\\	& \leq & 
	\sum_i \lambda_i \lpm E_f(U_{AB} |\psi_i\>) - E_f(|\psi_i\>) \rpm 
	\leq \max_i \lpm E_f(U_{AB} |\psi_i\>) - E_f(|\psi_i\>) \rpm
\label{eq:ef}
\,.
\eea
The second inequality is obtained by applying convexity of $E_f$ to the
first term, and the definition of the optimal decomposition in the
second term.
Thus, $E_{f,U}^{(1,*,{\rm cc})}-\epsilon$ can be attained on a pure
input state. \qed

{\bf Lemma 4:} 
For any $\epsilon > 0$, $E_{c,U}^{(1,*,{\rm cc})}-\epsilon$ can be
attained on a pure input state.

{\bf Proof:} Let $\rho$ attain the supremum in \eq{e1} up to
$\epsilon/2$.  That is, 
\bea
	E_{c,U}^{(1,*,{\rm cc})} - {\epsilon \over 2} & \leq & 
	E_c(U_{AB} \; \rho \; U_{AB}^\dagger) 
	- E_c(\rho) \,.
\label{eq:trivial}
\eea
For any $\delta > 0$, $\exists~m$ such that ${1 \over m} E_f(\rho^{\ot
m}) - E_c(\rho) \leq \delta$ \cite{Hayden00}.  
Substitute this into \eq{trivial} with $\delta = \epsilon/2$, 
\bea
	E_{c,U}^{(1,*,{\rm cc})} - {\epsilon \over 2} & \leq & 
	E_c(U_{AB} \; \rho \; U_{AB}^\dagger) 
	- {1 \over m} E_f(\rho^{\ot m}) + {\epsilon \over 2} \,.
\label{eq:trivial2}
\eea
Using weak additivity of $E_c$ and the fact $E_f \geq E_c$, 
the first term in the RHS of \eq{trivial2} can be rewritten: 
\bea
	E_{c,U}^{(1,*,{\rm cc})} - \epsilon & \leq & 
	{1 \over m} \lbm 
	E_f(U_{AB}^{\ot m} \rho^{\ot m} U_{AB}^{\dagger \ot m}) 
	- E_f(\rho^{m}) \rbm \,.
\eea
But the expression in the bracket represents the entanglement of
formation generated by a certain $m$-use protocol, and is no greater
than $m E_{f,U}^{(1,*,{\rm cc})}$ by the 1st half of Theorem 1.
Together with Lemma 3,
\bea
	E_{c,U}^{(1,*,{\rm cc})} - \epsilon & \leq & 
	\sup_{|\psi\>} \; E_f(U_{AB} |\psi\>) - E_f(|\psi\>) \,.
\eea
Finally, we replace $E_f$ by $E_c$ on the RHS since they coincide 
on pure states, 
\bea
	E_{c,U}^{(1,*,{\rm cc})} - \epsilon & \leq & 
	\sup_{|\psi\>} \; E_c(U_{AB} |\psi\>) - E_c(|\psi\>) \,,  
\eea
which proves our claim.  \qed

When Theorem 1 holds, lemmas about $E_U^{(1,*,{\rm cc})}$ hold for
$E_U^{(*,{\rm cc})}$ as well.  This fact, together with Lemmas 2-4,
lead to many useful corollaries.  These are given with numbers
matching those of the corresponding lemmas.

{\bf Corollary 2:} If Theorem 1 holds for $E$, then $E_U^{(*,{\rm
cc})}$ is achievable using the $1$-use protocol in
Fig.~(\ref{eq:egen}) with sufficiently large dimensional $A'$, $B'$.

{\bf Corollary 3.1:} Since $E_f$ is strongly additive on pure states, 
Lemma 3 implies that Theorem 1 holds for $E_f$.

{\bf Corollary 3.2, 4.1:} Corollary 3.1 and Lemma 4 imply that
$\forall_{\epsilon > 0}$, $E_{f,U}^{(*,{\rm cc})} - \epsilon$ and
$E_{c,U}^{(*,{\rm cc})} - \epsilon$ are each attainable with the
$1$-shot protocol in Fig.~(\ref{eq:egen}) on a pure input state.

{\bf Corollary 3$+$4:} $E_{f,U}^{(1,*,{\rm cc})} =
E_{c,U}^{(1,*,{\rm cc})}$ with common pure optimal inputs.  Same for
$E_{f,U}^{(*,{\rm cc})} = E_{c,U}^{(*,{\rm cc})}$.

It is unclear whether $E_{d,U}^{(1,*,{\rm cc})} - \epsilon$ can be
attained on a pure state.  Convexity in $E_{f,c}$ is required in our
proofs of Lemmas 3 and 4, but unlike $E_f$ and $E_c$, $E_d$ may not be
convex~\cite{edmono}.

Note that Theorem 1 is concerned with weak additivity of the
entanglement capacity of bidirectional channels, i.e., the protocol
uses only one type of nonlocal gate.  We can consider strong
additivity when different types of nonlocal gates are available:

{\bf Theorem 1S (1st half):} 
For a protocol with $n_i$ uses of the gate $U_i$, the maximum amount
of entanglement generated (given $*$, {\rm cc}) is no more than
$\sum_i n_i E_{U_i}^{(1,*)}$.

{\bf Theorem 1S (2nd half):} 
If $E$ is strongly additive or subadditive on the optimal input 
and strongly additive or superadditive on the optimal output
for each $U_i$, then repeating $n_i$ times the $1$-shot protocol for
$U_i$ for each $i$ generates an amount of entanglement no less than 
$\sum_i n_i E_{U_i}^{(1,*,{\rm cc})}$.

In particular, Theorem 1S holds for $E = E_{f,c}$, and the entanglement 
capacities are strongly additive given $*$ and {\rm cc}.

\subsection{Auxiliary resources are unnecessary when $E = E_c$ or $E_f$}
\label{sec:asymres}

In this subsection, we show that the resource $*$ is unnecessary in the
optimal asymptotic protocol in the previous subsection (repeating the
optimal $1$-shot protocol) for the specific measures $E_{\rm in} =
E_{\rm out} = E_{c,f}$.

By Lemma 4, when $E = E_c$, the optimal input and output of the
$1$-shot protocol are pure.
The amount of entanglement $E_{c,U}^{(*,{\rm cc})} - \epsilon =
E_{c,U}^{(1,*,{\rm cc})} - \epsilon$ can be generated by adapting an
argument in \cite{Dur00}.
The protocol first creates $m$ copies of the pure optimal input
$|\psi\>_{AA'BB'}^{\ot m}$ (inefficiently), and then repeats the
cycle: (1) apply $U^{\ot m}$, (2) ``concentrate''~\cite{Bennett95} the
outputs to EPR pairs and (3) dilute some of the EPR pairs to form
$|\psi\>_{AA'BB'}^{\ot m}$~\cite{Lo99a,Lo00rsp,Hayden02a,Harrow02a}.
For large $m$, dilution and concentration take $O(\sqrt{m})$ classical
communication and waste $O(\sqrt{m})$ amount of
entanglement~\cite{Hayden02a,Harrow02a}, both can be supplied by an
additional $O(\sqrt{m})$ uses of $U$ -- negligible for sufficiently
large $m$.
The cost of creating the first $m$ copies of the pure optimal input
inefficiently is also negligible when the cycle is repeated
sufficiently many times.
The same argument holds for $E = E_f$.  

{\bf Corollary 3.3, 4.2:} $E_{c,U}^{(*,{\rm cc})} =
E_{c,U}^{(\emptyset)}$ and $E_{f,U}^{(*,{\rm cc})} =
E_{f,U}^{(\emptyset)}$.  

The asymptotic entanglement capacity for $E=E_{c,f}$ under the most
general setting in \sec{edefs} can be generated with no initial
entanglement and without $*$ nor {\rm cc}.  The core part of the
optimal protocol is basically $1$-shot -- tensor product of the
optimal $1$-shot protocol.  The only collective steps, entanglement
concentration and dilution, are auxiliary.

Since no initial entanglement is required for the optimal asymptotic
generation of $E_f$ and $E_c$, one can relate the asymptotic
entanglement capacities to the {\em Schmidt number} of $U$.  Any
bipartite pure state can be written as $\sum_i \lambda_i |\phi_i\>_A
|\eta_i\>_B$ where $\lambda_i > 0$, $\sum_i \lambda_i^2 = 1$ and
$\{|\phi_i\>\}$, $\{|\eta_i\>\}$ are orthonormal sets of states.
Likewise, any $d \times d$ bipartite unitary gate can be written as
$\sum_i \lambda_i V_{iA} \otimes W_{iB}$ where $\lambda_i > 0$,
$\sum_i \lambda_i^2 = d^2$ and $\{V_{iA}\}$, $\{W_{iB}\}$ are sets of
operators orthonormal under the trace norm.
The Schmidt number of the bipartite pure state or
gate~\cite{Nielsen00,Nielsen98}, denoted as Sch$(\cdot)$, is the
unique number of terms in the above ``Schmidt decomposition.''  The
$\lambda_i$ are called the Schmidt coefficients.
We will repeatedly use the fact that the Schmidt number of a state is
nonincreasing under LOCC and that Sch$(U|\psi\>) \leq$ Sch$(U)$
Sch$(|\psi\>)$ (see Ch.~6.4.2 of~\cite{Nielsen98}).

{\bf Corollary 3.4, 4.3:} $E_{c,U}^{(*,{\rm cc})} = E_{f,U}^{(*,{\rm
cc})} \leq \log$$\;{\rm Sch}(U)$. 

{\bf Proof:} Since $E_{c,U}^{(*,{\rm cc})}$ and $E_{f,U}^{(*,{\rm
cc})}$ can be achieved without initial entanglement, the initial state
has Schmidt number 1, and the final state of a $t$-use protocol has
Schmidt number $\leq {\rm Sch}(U)^t$.  Hence, the output entropy of
entanglement is $\leq t \ss \log \ss {\rm Sch}(U)$, and
$E_{c,U}^{(*,{\rm cc})} = E_{f,U}^{(*,{\rm cc})} \leq \log \; {\rm
Sch}(U)$. \qed

{\bf Corollary 3.5, 4.4:} $E_{c,U}^{(1,\emptyset)} =
E_{f,U}^{(1,\emptyset)} \geq - \sum_i \lambda_i^2 \log \lambda_i^2$ where
$\{\lambda_i\}$ are the Schmidt coefficients of $U$. 

{\bf Proof:} This is the entanglement generated when $|\psi\>_{AA'BB'} =
|\Phi_d\>_{AA'} |\Phi_d\>_{BB'}$ in \eq{e1}~\cite{Terhal01u}.  \qed

Interested readers can repeat the above analysis for other measures.  
It holds for $E=E_d$ if the optimal input $\rho$ is pure or if $\rho$
satisfies $E_c(\rho) = E_d(\rho)$ (by replacing concentration with
distillation of the optimal output and replenishing the optimal input
$\rho$ using $E_d(\rho)$ EPR pair per copy of $\rho$ and classical
communication (see Appendix~\ref{sec:ccdist})).

\subsection{Different input and output entanglement measures}
\label{sec:diffmeas}

Each choice of entanglement measures for the input and output can 
be given an operational meaning.  
We consider the important example of creating EPR pairs in this
subsection, which requires different entanglement measures for the
input and output.
Alice and Bob fabricate the possibly mixed optimal input state and
distill entanglement from the output.
Thus, the appropriate choices for the input and output entanglement
measures are the entanglement cost $E_c$ and the distillable
entanglement $E_d$.\footnote{Note that $t$ is finite, but we have
chosen the asymptotic measures $E_c$ and $E_d$.  We are mainly
interested in protocols with large $t$, with the understanding that
the $1$-shot capacity $E_{c \ra d, U}^{(1,*)}$ is achieved with
collective pre- and post-processing.}
Let ${\cal P}_t$ denote an optimal $t$-shot protocol and the
corresponding quantum operation, and let $\rho$ be the optimal input
to within $\epsilon$ (again we omit the system label $AA'BB'$).
Then
\bea
	E_{c \ra d, U}^{(t,*,{\rm cc})} - \epsilon 
	& \leq & {1 \over t} \lbm E_d({\cal P}_t(\rho)) - E_c(\rho) \rbm 
	~~ \leq ~~ {1 \over t} \lbm E_c({\cal P}_t(\rho)) - E_c(\rho) \rbm 
\non 
\\
	& \leq & \sup_{|\psi\>}~  
		E_c(U_{AB} |\psi\>) - E_c(|\psi\>)
\label{eq:usecor4}
\\
	& = & \sup_{|\psi\>}~ E_d(U_{AB} |\psi\>) 
					- E_c(|\psi\>)  
\non
\\
	& = & E_{c,U}^{(1,*)} \leq E_{c \ra d, U}^{(1,*)}
\non
\eea
where we have used Corollary 4 in Sec.~\ref{sec:edd} to obtain
\eq{usecor4}.  
This means that the asymptotic capacity to create EPR pairs is 
\bea
	E_{c \ra d, U}^{(t,*,{\rm cc})} ~ = ~\sup_{|\psi\>}~ 
	E_c(U_{AB} |\psi\>) - E_c(|\psi\>) 
\eea
and the protocol in \sec{asymres} is optimal for creating EPR pairs 
even in the most general setting described in \sec{edefs}. 

Furthermore, since the optimal output is pure, and $E_d$, $E_c$ are
strongly additive on pure states strong additivity (Theorem 1S) holds
when different types of gates are given.

\subsection{Summary}
\label{sec:esum}

We summarize our results obtained so far: 
\begin{enumerate}
\item $E_U^{(1,*)} = E_U^{(1,*,{\rm cc})} \geq E_U^{(*,{\rm cc})}$ 
	for all $E$.
\item $E_U^{(1,*)} = E_U^{(*,{\rm cc})} = E_U^{(*)}$
	for all $E$ weakly additive on the $1$-shot optimal 
	input and output.  
\item $E_U^{(1,*)} = E_U^{(\emptyset)}$ for $E=E_{c,f}$.
\item $E_{c,U}^{(1,*)} = E_{c \ra d,U}^{(*,{\rm cc})}$.
\end{enumerate}

In particular, when $E_{\rm in} = E_{\rm out} = E_f$, or when $E_{\rm
in} = E_c$, and $E_{\rm out} = E_c$ or $E_d$, the asymptotic
capacities become independent of the availability of $*$ and {\rm cc},
and they are all equal to $E^{(1,*)}_{e, U} = \Delta E_U$ in
\eq{eentropy}.
The only capacity mentioned above that is different from $\Delta E_U$
is $E^{(1,\emptyset)}_{\inout, U}$.
We will study these two capacities in \sec{gbdd} and \sec{evilgates}
in more detail.  

As an aside, when $E=E_c$ (or $E_f$), if $E^{(1,\emptyset)}_{c,U} <
E^{(\emptyset)}_{c,U}$, then $E^{(t,\emptyset)}_{c,U} <
E^{(\emptyset)}_{c,U}$ for all finite $t$.
This is because $t E_{c,U}^{(t,\emptyset)} \leq
E^{(1,\emptyset)}_{c,U} + (t-1) E_{c,U}^{(t-1,*)} =
E^{(1,\emptyset)}_{c,U} + (t-1) E_{c,U}^{(\emptyset)} < t
E_{c,U}^{(\emptyset)}$.

\section{Classical capacities of bidirectional channels} 
\label{sec:def}

If Alice and Bob have access to a nonlocal gate $U$ to couple their
systems, then the classical communication capacity of $U$ is the
maximum asymptotic number of classical bits that can be reliably
transmitted per use of $U$.
%
Communication can be achieved simultaneously in both directions,
with possible tradeoffs. 
Free local resources as stated in \sec{structure} and shared
classical randomness are always allowed.

In the context of classical communication, the most important
auxiliary resource is free entanglement.  Communication is called
``assisted/unassisted'' when the resource is/not available.  

The most general protocol (see \sec{bichan}) with $t$ uses of $U$ can be
represented as:
%
\vspace*{-2ex}
\bea
\label{eq:ccgen}
\centering
\setlength{\unitlength}{0.6mm}
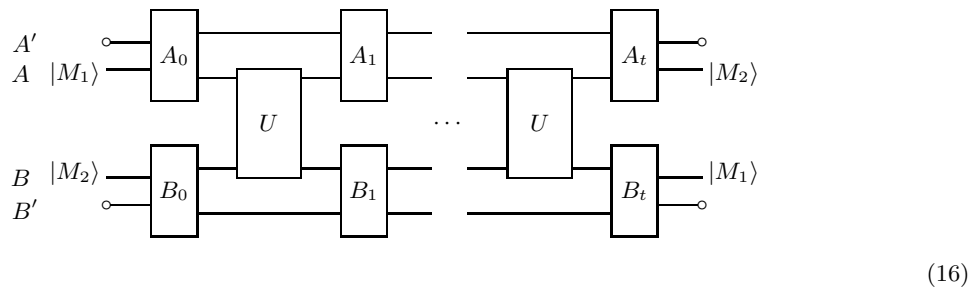
\begin{picture}(185,70)
{\small
\put(7,48){\makebox(10,10){$A'$}}
\put(7,41){\makebox(8,10){$A$}}
\put(7,18){\makebox(8,10){$B$}}
\put(7,11){\makebox(10,10){$B'$}}
}
\put(16,19){\makebox(14,10){$|M_2\>$}}
\put(30,17){\circle{2}}
\put(31,17){\line(1,0){9}}
\put(30,23){\line(1,0){10}}
\put(40,10){\framebox(10,20){$B_0$}}
\put(16,41){\makebox(14,10){$|M_1\>$}}
\put(30,47){\line(1,0){10}}
\put(31,53){\line(1,0){9}}
\put(30,53){\circle{2}}
\put(40,40){\framebox(10,20){$A_0$}}
\put(59,23){\framebox(14,24){$U$}}
\put(50,15){\line(1,0){32}}
\put(50,25){\line(1,0){9}}
\put(73,25){\line(1,0){9}}
\put(82,10){\framebox(10,20){$B_1$}}
\put(50,45){\line(1,0){9}}
\put(73,45){\line(1,0){9}}
\put(50,55){\line(1,0){32}}
\put(82,40){\framebox(10,20){$A_1$}}
\put(92,15){\line(1,0){10}}
\put(92,25){\line(1,0){10}}
\put(92,45){\line(1,0){10}}
\put(92,55){\line(1,0){10}}
\put(101,30){\makebox(10,10){$\cdots$}}
\put(110,15){\line(1,0){32}}
\put(110,25){\line(1,0){9}}
\put(110,45){\line(1,0){9}}
\put(110,55){\line(1,0){32}}
\put(133,25){\line(1,0){9}}
\put(133,45){\line(1,0){9}}
\put(119,23){\framebox(14,24){$U$}}
\put(142,10){\framebox(10,20){$B_t$}}
\put(142,40){\framebox(10,20){$A_t$}}
\put(162,19){\makebox(14,10){$|M_1\>$}}
\put(162,17){\circle{2}}
\put(152,17){\line(1,0){9}}
\put(152,23){\line(1,0){10}}
\put(162,41){\makebox(14,10){$|M_2\>$}}
\put(152,47){\line(1,0){10}}
\put(152,53){\line(1,0){9}}
\put(162,53){\circle{2}}
\end{picture}
\eea
\vspace*{-6ex}

In Fig.~(\ref{eq:ccgen}), $A$ and $B$ label the systems acted on by
$U$ during the protocol and the systems carrying the classical
messages before and after the protocol, while $A'$ and $B'$ label the
rest of Alice and Bob's systems.  The dimensions of $A$ and $B$ are
converted to $d$ by the initial operation $A_0 \ot B_0$ and are
further converted by the final operation $A_t \ot B_t$.  The freedom
to apply $U$ to any register is included as swap operations in $A_i$
and $B_i$.
Without loss of generality, the local operations $A_j$, $B_j$ can be
assumed unitary, since measurements can be deferred until the end of
the protocol.
In fact, no measurements are needed, except for the final readout of
the transmitted messages.
In the unassisted case, the initial ancilla state $|\psi_{\rm
anc}\>_{A'B'}$ can be taken to be $|0r\>_{A'}|0r\>_{B'}$ with random
$r$, which can generate an arbitrary separable state.  In the assisted
case, $|\psi_{\rm anc}\>_{A'B'}$ can be taken to be the maximally
entangled state, which can generate an arbitrary entangled state.

Each protocol allows a certain amount of forward and backward
classical communication, giving a pair of achievable rates
$(R_\ra,R_\la)$ for the gate.
\begin{quote}
\begin{definition} 
{\rm 
A pair of rates $(R_\ra,R_\la)$ is said to be {\em achievable} by a
gate $U$ if it is possible to intersperse $t$ uses of $U$ with
local unitaries $A_j \ot B_j$, such that an $n_1$-bit message $M_1$
from Alice to Bob and an $n_2$-bit message $M_2$ from Bob to Alice are
communicated with high fidelity, and $R_\ra \geq n_1/t$, $R_\la
\geq n_2/t$.  Mathematically $(R_\ra,R_\la)$ is achievable if:
\bea
\hspace*{-3ex} & & 
\forall \, \epsilon \,,
~\exists \, t \,, 
~\exists \, n_1 \!\!\geq \! tR_\ra \,, 
~\exists \, n_2 \!\! \geq \! tR_\la \,,
~\exists \{ A_j \ot B_j\}_{j=0}^t \,,
\non
\\
\hspace*{-3ex} &  {\rm ~s.t.}  & 
\forall M_1 \in \{0,1\}^{n_1} \,, ~\forall M_2 \in \{0,1\}^{n_2} \,,
\non
\\
& & |\psi_{\rm out} \>= ( A_t \otimes B_t ) \; U \; \cdots \; 
U \; (A_1 \otimes B_1) \; U \; (A_0 \otimes B_0) \;  
(|M_1\>_A \, |M_2\>_B \, |\psi_{\rm anc}\>_{A'B'}) 
\non
\\
\hspace*{-3ex} & {\rm and} & F \lpm |M_2\>_A \ot |M_1\>_B \; , 
\; \tr_{A'B'} \, |\psi_{\rm out}\> \<\psi_{\rm out}| \rpm 
> 1-\epsilon 
\label{eq:C_U-def}
\eea
}
\end{definition}
\end{quote}

In the above definition, the fidelity $F$ between two states
$|\psi\>\<\psi|$ and $\rho$ is given by $\<\psi|\rho|\psi\>$ (this is a
simplified expression when one of the states is pure).

We first discuss unassisted capacities, and the assisted capacities 
are defined in exactly the same way.  
Each gate $U$ defines a region of achievable unassisted rate-pairs
$(R_\ra,R_\la)$.
The region is convex by using mixed strategies.
Furthermore, if $(R_\ra,R_\la)$ is achievable, so is any
$(R_\ra',R_\la')$ where $R_\ra' \leq R_\ra$ and $R_\la' \leq R_\la$.
In particular, the boundary of the achievable region never has
positive slope (see Fig.~(\ref{eq:ratespic})).
Thus, the forward and backward capacities can always be achieved at
the boundary points, and can be defined respectively as
\bea 
	C_{\ra,U} &=& 
	\sup\{R : (R,0) \mbox{ is achievable by } U\}
\non
\\
	C_{\la,U} &=& 
	\sup\{R : (0,R) \mbox{ is achievable by } U\}
\non
\eea
We can also define various bidirectional capacities, for example, the 
duplex and the total capacities:
\bea 
	C_{\lra,U} &=& 
	\sup\{R : (R,R) \mbox{ is achievable by } U\}
\non
\\
	C_{+,U} &=& 
	\sup\{R_\la+R_\ra : (R_\la,R_\ra) \mbox{ is achievable by } U\}
\non
\eea
We omit the subscript $U$ when the notation is too cumbersome.  The
following is a schematic diagram for the achievable region and the
definitions of the various capacities.  We present all the {\em known}
properties and intentionally show the features that are not ruled out,
such as the asymmetry of the region, and the nonzero curvature of the
boundary.
%
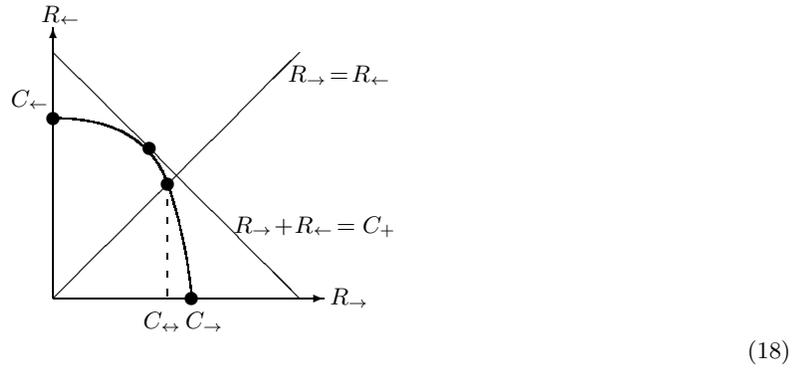
\bea
\centering
\setlength{\unitlength}{0.8mm}
\begin{picture}(72,62)
\put(10,10){\vector(1,0){45}}
\put(10,10){\vector(0,1){45}}
\put(10,10){\line(1,1){41}}
\put(10,51){\line(1,-1){41}}
\qbezier(10,40)(25,40)(29,29)
\qbezier(29,29)(32,20)(33,10)
\put(26,35){\circle*{2}}
\put(29,29){\circle*{2}}
\put(33,10){\circle*{2}}
\put(10,40){\circle*{2}}
\put(56,9){\makebox{$R_\ra$}}
\put(8,56){\makebox{$R_\la$}}
\put(25,5){\makebox{$C_\lra$}}
\multiput(29,26.5)(0,-3){6}{\line(0,-1){1}}
\put(49,46){\makebox{$R_\ra \! = \! R_\la $}}
\put(40,21){\makebox{$R_\ra \! + \! R_\la \! = C_+$}}
\put(32,5){\makebox{$C_\ra$}}
\put(3,42){\makebox{$C_\la$}}
\end{picture}
\label{eq:ratespic}
\eea
%
There are much simpler examples -- the unassisted achievable region
for {\sc cnot} and {\sc swap} are similar triangles with vertices
$\{(0,0), (0,1), (1,0)\}$ and $\{(0,0), (0,2), (2,0)\}$ respectively
(see \sec{evilgates}).

In general, little is known about the unassisted achievable region of
$(R_\ra,R_\la)$ besides the convexity and the monotonicity of its
boundary.
The most perplexing question is perhaps whether the region has
reflective symmetry about $R_\ra = R_\la$, which implies $C_\ra =
C_\la$ and $C_+ = 2 C_\lra$.
Refs.~\cite{Dur00,Kraus00} show that any two-qubit gate or Hamiltonian
is locally equivalent to one with Alice and Bob interchanged, so that
the achievable region is indeed symmetric.
This implies the conjecture in \cite{Eisert00} that the {\em one-shot}
forward and backward unassisted capacities are equal.
In higher dimensions, \cite{Bennett01} shows that there are
Hamiltonians (and so unitary gates) that are intrinsically asymmetric.
However, it remains open whether the achievable rate pairs are
symmetric, or more weakly, whether $C_\ra=C_\la$ or $C_+=2C_\lra$.

Assisted capacities $C^{E}_{\ra,U}$, $C^{E}_{\la,U}$,
$C^{E}_{\lra,U}$, $C^{E}_{+,U}$ can be defined in exactly the same
manner, now the ancilla $|\psi_{\rm anc}\>_{A'B'}$ is maximally
entangled instead of being $|0r\>_{A'}|0r\>_{B'}$ in the definition of
the achievable rate pairs in \eq{C_U-def}.  The properties and open
questions of the achievable region are also the same as those in the
unassisted case.  Two simple examples are the assisted achievable
regions for {\sc cnot} and {\sc swap}, they are similar squares with
vertices $\{(0,0), (0,1), (1,0), (1,1)\}$ and $\{(0,0), (0,2), (2,0),
(2,2)\}$ respectively (see \sec{evilgates}).

Entanglement assistance greatly simplifies the analysis of the
classical capacities $C_E$ of the {\em usual} (unidirectional) quantum
channels.\footnote{Note that the resource $E$ appears as a subscript
of the capacity for a unidirectional channel, and as a superscript for
a bidirectional channel. }
An expression for $C_E$ has been found and proved to be strongly
additive~\cite{Shor01,Holevo01}.
The study of $C_E$ also provides useful upper bounds for the
unassisted capacities and insights to the classification of
channels~\cite{qrst}.
In the next section, we derive a simple expression for
$C^{E}_{\ra,U}$ and $C^{E}_{\la,U}$, the $1$-way (forward or
backward) {\em entanglement-assisted} capacity of any bidirectional
channel.
Surprisingly, this capacity is also strongly additive, as in 
the unidirectional case!

Comparison of the two problems of generating entanglement and
classical communication will be given in \sec{diss}, and the two
resulting capacities are related in Sec.~\ref{sec:gbdd}.

\section{Entanglement-assisted one-way classical capacity} 
\label{sec:ce}

\subsection{Preliminaries and definitions} 

In this section we derive expressions for $C^{E}_{\ra,U}$ and
$C^{E}_{\la,U}$, as defined in \eq{C_U-def} with $|\psi_{\rm anc}\>$
being a maximally entangled state.
Without loss of generality, we focus on $C^{E}_{\ra,U}$.
It can be evaluated using the general framework of $1$-way classical
communication with quantum
resources~\cite{Hausladen96,Holevo98,Schumacher97}.
%
%
In this framework, suppose classical messages $i$, occurring with
probabilities $p_i$, are encoded in the ``signal states'' $\eta_i$
received by Bob, forming an ensemble ${\cal E} = \{p_i, \eta_i\}$.
The information on $i$ obtained by measuring a signal state is upper
bounded by the {\em Holevo information} \tchi for the ensemble $\cal
E$, defined as
\bea
	\mchi \lpm {\cal E} \! = \! \{p_i,\eta_i\}  \rpm 
	\equiv 
 	S \lpL \sum_i p_i \eta_i \rpL - \sum_i p_i S (\eta_i) 
\,.
\label{eq:chidef}
\eea
The Holevo-Schumacher-Westmoreland (HSW) Theorem states that this
amount of mutual information per signal state is achievable given the
ability to transmit an asymptotically large number of signal states.
(See \cite{Holevo98,Schumacher97} and Ch. 12.3.2 of \cite{Nielsen00}.) 

%
We will see that the optimal methods to generate EPR pairs (see
Secs.~\ref{sec:asymres}-\ref{sec:diffmeas}) and entanglement-assisted
classical communication have many similarities.
The respective goals are to maximize the increase in entanglement and
the Holevo information.
The optimal asymptotic strategies in both cases are to repeat the
$1$-shot protocol, with an optimal input state in the former and with
an optimal input ensemble in the latter.
In the case of entanglement generation, allowing the most general
$1$-shot optimal input with arbitrary ancillas and initial
entanglement makes the $1$-shot capacity equal to the asymptotic ones.
Likewise, we will allow the most general $1$-shot input ensemble for
assisted classical communication, and will show that the resulting
$1$-shot capacity is equal to the asymptotic capacities by
establishing a method to ``replenish'' the optimal input ensemble
(analogous to concentration and dilution in entanglement generation).

Let ${\cal E} = \{p_i, |\psi_i\>_{A\!A'\!B\!B'}\}$ be an ensemble of
bipartite states.
A trace-preserving operation acts on ${\cal E}$ by acting on each
component state (preserving its probability).
For example, we will write 
$U {\cal E} = \{p_i, U_{A\!B} |\psi_i\>_{A\!A'\!B\!B'}\}$, 
$\tr_{A\!A'} {\cal E} = \{p_i, \tr_{A\!A'} 
		|\psi_i\>\<\psi_i|_{A\!A'\!B\!B'}\}$, and 
$\tr_{A\!A'} U {\cal E} = \{p_i, \tr_{A\!A'}
U_{A\!B} |\psi_i\>\<\psi_i|_{A\!A'\!B\!B'} U_{A\!B}^\dagger \}$.
We have the following definitions analogous to those in \sec{edefs}: 
\begin{quote}
\begin{definition} 
{\rm The $t$-shot Holevo information capacity of $U$ is the maximum
increase in Holevo information per use of $U$ due to any protocol
${\cal P}_t$ that uses $U$ $t$ times, the auxiliary resources labelled
$r$, and the local resources specified in Sec.~\ref{sec:structure}.
There are two possible $t$-shot capacities, depending on the allowed 
input ensembles: \\
1. when the input ensemble ${\cal E}_0$ is restricted to satisfy 
$\mchi(\tr_{A\!A'} {\cal E}_0) = 0$
\bea
	\Delta \mchi_{\ra,U}^{(t,\emptyset,r)} \equiv 
	\sup_{{\cal P}_t} \; \sup_{{\cal E}_0} \; {1 \over t} \ss  
	\mchi(\tr_{A\!A'} {\cal P}_t \ss {\cal E}_0)  
\label{eq:dchideft0}
\eea
2. when the input ensemble $\cal E$ is unrestricted:  
\bea
	\Delta \mchi_{\ra,U}^{(t,*,r)} \equiv 
	\sup_{{\cal P}_t} \; \sup_{{\cal E}} \; {1 \over t} \lbm 
	\mchi(\tr_{A\!A'} {\cal P}_t \ss {\cal E}) -
	\mchi(\tr_{A\!A'} {\cal E}) \rbm 
\label{eq:dchideft+}
\eea
}
\end{definition}
\end{quote}
Since we always assume free entanglement as an auxiliary resource, and
we always focus on forward capacity, we omit $r=E$ and $\ra$ in the 
above notation: 
\bea
	\Delta \mchi_{U}^{(t,\emptyset)} \equiv 
	\Delta \mchi_{\ra,U}^{(t,\emptyset,E)} \,,~~~{\rm and}~~~
 	\Delta \mchi_{U}^{(t,*)} \equiv 
	\Delta \mchi_{\ra,U}^{(t,*,E)} 
\eea
We have 
\bea
	C_{\ra,U}^{E} 
	= \sup_t \; \Delta \mchi_{U}^{(t,\emptyset)} 
\label{eq:ceexp}
\eea
Note that it is unnecessary to consider mixed state ensembles in
\eq{dchideft0} and \eq{dchideft+} -- we can replace a mixed state
$\rho_{AA'BB'}$ by its purification $|\psi\>_{A\!A'\!A''\!B\!B'}$,
where $A''$ is the purifying system, without affecting
$\tr_{A\!A'\!A''} {\cal E}$ nor $\tr_{A\!A'\!A''} {\cal P}_t {\cal
E}$.

In the next two subsections, we will prove $C_{\ra,U}^{E} = \Delta
\mchi_{U}^{(1,*)}$.
We first prove that $C_{\ra,U}^{E} \leq \Delta \mchi_{U}^{(1,*)}$, 
and then we describe a protocol to achieve the upper bound, thereby
proving additivity and providing an optimal asymptotic strategy.
%

\subsection{An additive upper bound} 
\label{sec:chiubdd}

We first prove an analogue of Lemma 1:

{\bf Lemma 5:}
$C^{E}_{\ra,U} \leq 2 \log d$ and 
$C^{E}_{\la,U} \leq 2 \log d$. 

{\bf Proof:} Consider a $t$-use protocol.  Replace each use of $U$ by
double teleportation (see Lemma 1).  If the original protocol consumes
and produces $C_{\rm in}$ and $C_{\rm out}$ bits of forward
communication, the modified protocol consumes and produces $C_{\rm in}
+ 2 \, t \log d$ and $C_{\rm out}$ bits of forward communication.  By
causality~\cite{Holevo73d} of the modified protocol $2 \, t \log d
\geq C_{\rm out} - C_{\rm in}$.  Hence $C^{E}_{\ra,U} \leq 2 \log d$.
Similarly $C^{E}_{\la,U} \leq 2 \log d$.  (Note that the above proof
is stronger than we need, since we have allowed $C_{\rm in} \neq 0$.)
\qed

Consider the best $1$-shot protocol to increase the Holevo
information.
Since local operations do not increase mutual information, the optimal
$1$-shot protocol is to just apply $U$, as in Fig.~(\ref{eq:egen}).
Thus  
\bea
	\Delta \mchi_U^{(1,*)} = \sup_{\cal E} \lbm 
		\mchi(\tr_{A\!A'}U{\cal E}) 
		- \mchi(\tr_{A\!A'} {\cal E}) \rbm 
\label{eq:dchi}
\eea
where the supremum is over the most general bipartite pure state
ensemble ${\cal E} = \{p_i, |\psi_i\>_{A\!A'\!B\!B'}\}$.  

We now consider the asymptotic problem.
Using the same idea that proves Theorem~1 (1st half), we obtain the
following analogue.

{\bf Theorem 2 (1st half):} $\Delta \mchi_U^{(t,*)} \leq \Delta
\mchi_U^{(1,*)}$.

{\bf Proof:} Consider the most general protocol ${\cal P}_t$ with $t$
uses of $U$ (such as depicted in Fig.~(\ref{eq:ccgen})).  Let $\cal E$
be an arbitrary bipartite input ensemble.
Then, the total increase in \tchi is upper bounded by the sum of the 
stepwise increases. 
Since local operations cannot increase \mchi, and the increase in
\tchi by each use of $U$ is bounded by \eq{dchi}, 
\bea 
	\mchi(\tr_{A\!A'} {\cal P}_t {\cal E}) - 
	\mchi(\tr_{A\!A'} {\cal E}) \leq 
	t \; \Delta \mchi_U^{(1,*)} 
\,,
\label{eq:thm21sthalf}
\eea
from which the Theorem is immediate.  \qed

It follows from \eq{thm21sthalf} and \eq{ceexp} that $C^E_\ra \leq
\Delta \mchi_U^{(1,*)}$.

\subsection{Protocol to achieve the upper bound of $C^E_{\ra,U}$} 
\label{sec:lglass}

%
In optimal asymptotic entanglement generation, the following basic cycle 
is repeated: \\
(1) convert EPR pairs into $n$ copies of the optimal input state, \\
(2) apply the gate to each, \\
(3) convert the $n$ copies of optimal output state into EPR pairs. \\
More EPR pairs are obtained in (3) than used in (1) -- as excess
entanglement generated. 

In entanglement-assisted classical communication, we want 
a similar basic cycle: \\
(1) convert classical communication to create $n$ states drawn
from the optimal {\em input} ensemble, \\
(2) apply the gate to each state, \\ 
(3) convert the states from the optimal {\em output} ensemble into 
classical communication. \\
Step (1) is called remote state
preparation~\cite{Lo00rsp,Bennett00,Devetak01} (RSP), a procedure
whereby Alice helps Bob to construct quantum states of her choice in
his laboratory using entanglement and classical communication.
In RSP, Alice performs a measurement on her half of the shared 
entangled state, sends the outcome to Bob, who conditioned on the
outcome operates on his half of the shared entangled state to complete
the RSP.
It is known \cite{peterrsp} how to approximately prepare $n$ pure
bipartite states from an ensemble ${\cal E}$ with free entanglement
and $n \ss \mchi(\tr_{AA'} {\cal E}) + o(n)$ bits of classical
communication. 
Step (3) follows from the HSW Theorem: Alice can communicate $\approx
n \ss \mchi(\tr_{AA'} U {\cal E}) - o(n)$ bits to Bob reliably if she
can prepare $n$ states in the output ensemble $U {\cal E}$.
Just like the case of generating entanglement, $n$ is chosen large
enough to ensure the efficiency of steps (1) and (3).

%
When describing and analyzing the protocol, we loosely call ${\cal E}$
the optimal ensemble achieving the supremum in \eq{dchi}.
For arbitrarily small $\epsilon$, ${\cal E}$ is chosen so
that $\mchi(\tr_{A\!A'}U{\cal E}) - \mchi(\tr_{A\!A'} {\cal E}) \geq
\Delta \mchi_U^{(1,*)} - \epsilon$.  Since how $\epsilon$ enters the
following analysis is obvious, and the analysis is independent of the
choice of $\epsilon$ and ${\cal E}$, $\epsilon$ is omitted for
simplicity.

{\bf Protocol that achieves $C^E_{\ra,U} = \Delta \mchi_U^{(1,*)}$:}  

Let ${\cal E}$ be the optimal ensemble.  
If Alice is given $n \, \mchi(\tr_{AA'} {\cal E})$ bits of classical
communication as an initial resource, she can transmit $k$ messages
$M_{i=1,\cdots,k}$ each of length $ n \, \Delta \mchi_U^{(1,*)}$ (a
total of $n \, k \, \Delta \mchi_U^{(1,*)}$ bits) with $nk$ uses of
$U$ as follows:
\begin{itemize}
\item {\em Alice's preprocessing:} ~Alice determines $k$ messages
$N_i$ each of $n \ss \mchi(\tr_{AA'} U {\cal E})$ bits.
Each $N_i$ has 2 parts: the message $M_i$ of length $ n \ss \Delta
\mchi_U^{(1,*)}$, and an RSP instruction $R_i$ of length $n
\mchi(\tr_{AA'} {\cal E})$ for Bob to create a state $|\phi_{i+1}\> \in
{\cal E}^{\ot n}$ such that $U^{\ot n} |\phi_{i+1}\> \in (U {\cal
E})^{\ot n}$ encodes $N_{i+1}$ (by the HSW Thm).
In order to generate $R_i$ for $N_i$, Alice needs to determine
$|\phi_{i+1}\>$ and to perform her measurement for the RSP of
$|\phi_{i+1}\>$.  This in turns requires knowledge of $N_{i+1}$.
So Alice first computes the last message $N_k$ (in which $M_k$ is
known and $R_k$ is irrelevant), classically calculates $|\phi_k\>$,
performs measurement for the RSP of $|\phi_k\>$ to find out $R_{k-1}$
in $N_{k-1}$, works her way {\em backwards} through $N_{k-1}$,
$\cdots$, $N_1$, determining $|\phi_i\>$ from $N_i$ and
performing measurement for RSP for $|\phi_i\>$ for decreasing $i$.

\item {\em Quantum protocol:} ~Alice uses the given initial classical
communication to create $|\phi_{1}\>$, which she shares with Bob.  Then
$U^{\ot n}$ is applied to convert it to $U^{\ot n}|\phi_1\>$, Bob
reads off the message $N_1$, which consists of $R_1$ to instruct him
to do RSP for $|\phi_2\>$ and so on.
\end{itemize}

The protocol is summarized in Fig.~(\ref{eq:lookingglass}). 

\vspace*{-7ex}
\bea 
\centering
\setlength{\unitlength}{0.6mm}
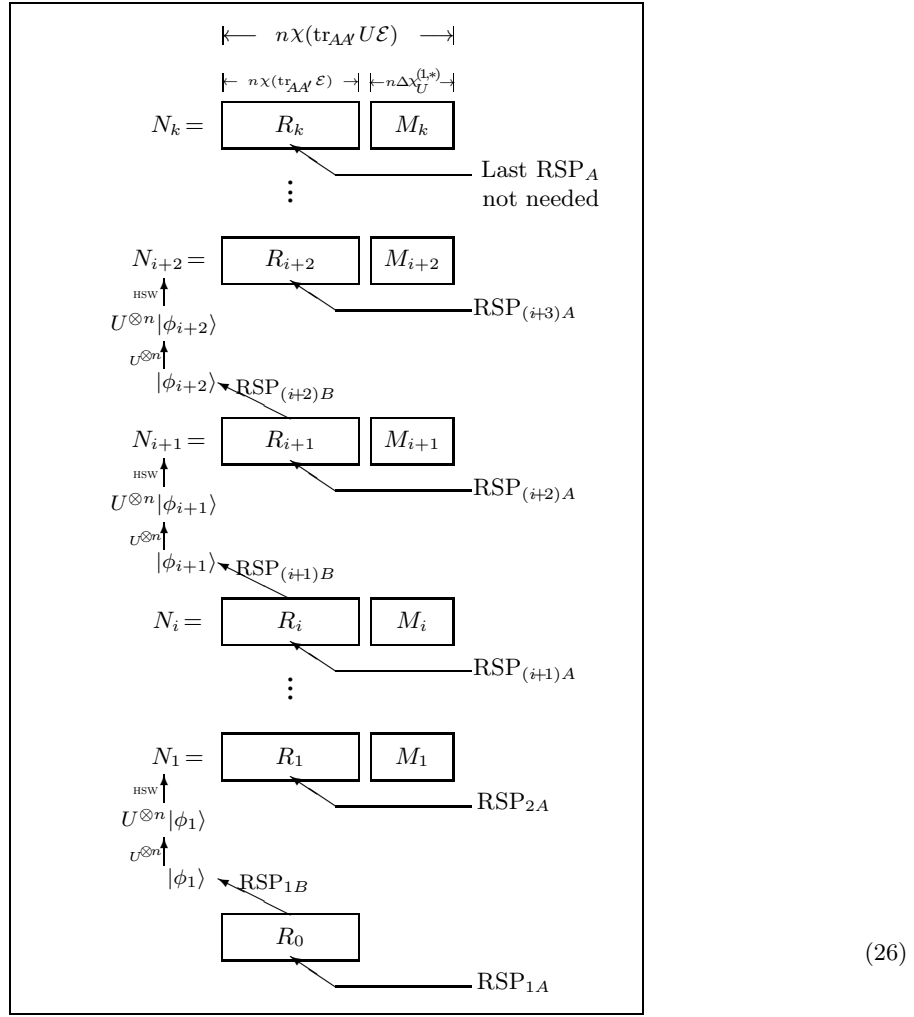
\begin{picture}(140,220)
\put(0,-12){\framebox(140,224){}}

{\footnotesize
\put(47.1,200){\makebox(51,10){$\longleftarrow ~ \,
n \ss \mchi(\tr_{\!A\!A\!'\!} \, U {\cal E}) ~ \; \longrightarrow$}}}

{\tiny
\put(47,189.2){\makebox(30,10){$\la \,
	n \mchi(\tr_{\!\!A\!A\!'}{\cal E}) \, \ra $}}
\put(80,190){\makebox(18,10){$\la\!\! 
	n \hspace*{-0.3ex} \Delta 
	\hspace*{-0.4ex} \mchi_{\!U}^{\!(\!1\!,\!*\!)}\! \!\!\ra$}}
}

\linethickness{0.01mm}
\put(79.7,192.5){\line(0,1){4}}
\put(98.1,192.5){\line(0,1){4}}

\put(47,192.5){\line(0,1){4}}
\put(77,192.5){\line(0,1){4}}

\put(47,203){\line(0,1){4}}
\put(98.1,203){\line(0,1){4}}

\linethickness{0.2mm}

\put(72,64){\vector(-3,2){10}}
\put(72,64){\line(1,0){30}}
\put(105,60){\makebox(18,8){RSP$_{(i\!+\!1)A}$}}

\put(72,34){\vector(-3,2){10}}
\put(72,34){\line(1,0){30}}
\put(105,30){\makebox(13,10){RSP$_{2A}$}}

\put(72,-6){\vector(-3,2){10}}
\put(72,-6){\line(1,0){30}}
\put(105,-10){\makebox(13,10){RSP$_{1A}$}}

\put(72,104){\vector(-3,2){10}}
\put(72,104){\line(1,0){30}}
\put(105,100){\makebox(18,8){RSP$_{(i\!+\!2)A}$}}

\put(72,144){\vector(-3,2){10}}
\put(72,144){\line(1,0){30}}
\put(105,140){\makebox(18,8){RSP$_{(i\!+\!3)A}$}}

\put(72,174){\vector(-3,2){10}}
\put(72,174){\line(1,0){30}}
\put(105,170){\makebox(25,10){Last RSP$_A$}}
\put(105,164){\makebox(25,10){not needed}}

{\footnotesize 
\put(52,12){\makebox(13,10){RSP$_{1B}$}}
\put(32,13){\makebox(14,10){$|\phi_1\>$}}
\put(27,26){\makebox(14,10){$U^{\ot n}|\phi_1\>$}}
\put(34,35){\vector(0,1){6}}
\put(62,10){\vector(-2,1){16}}
\put(34,21){\vector(0,1){6}}
}

{\tiny
\put(25,18.5){\makebox(10,10){$U^{\!\ot \! n}$}}
\put(25,33){\makebox(10,9){\sc hsw}}
}

{\footnotesize 
\put(52,82){\makebox(18,8){RSP$_{(i\!+\!1)B}$}}
\put(32,83){\makebox(14,10){$|\phi_{i+1}\>$}}
\put(27,96){\makebox(14,10){$U^{\ot n}|\phi_{i+1}\>$}}
\put(34,105){\vector(0,1){6}}
\put(62,80){\vector(-2,1){16}}
\put(34,91){\vector(0,1){6}}
}

{\tiny
\put(25,88.5){\makebox(10,10){$U^{\! \ot \! n}$}}
\put(25,103){\makebox(10,9){\sc hsw}}
}

{\footnotesize 
\put(52,122){\makebox(18,8){RSP$_{(i\!+\!2)B}$}}
\put(32,123){\makebox(14,10){$|\phi_{i+2}\>$}}
\put(27,136){\makebox(14,10){$U^{\ot n}|\phi_{i+2}\>$}}
\put(34,145){\vector(0,1){6}}
\put(62,120){\vector(-2,1){16}}
\put(34,131){\vector(0,1){6}}
}

{\tiny 
\put(25,128.5){\makebox(10,10){$U^{\! \ot \! n}$}}
\put(25,143){\makebox(10,9){\sc hsw}}
}

\put(30,180){\makebox(14,10){$N_k \! = $}}
\put(47,180){\framebox(30,10){$R_k $}}
\put(80,180){\framebox(18,10){$M_k $}}

\put(28,110){\makebox(14,10){$N_{i+1} \! = $}}
\put(47,110){\framebox(30,10){$R_{i+1} $}}
\put(80,110){\framebox(18,10){$M_{i+1} $}}

\put(28,150){\makebox(14,10){$N_{i+2} \! = $}}
\put(47,150){\framebox(30,10){$R_{i+2} $}}
\put(80,150){\framebox(18,10){$M_{i+2} $}}

{\Large
\put(47,167){\makebox(30,10){$\cdot$}}
\put(47,165){\makebox(30,10){$\cdot$}}
\put(47,163){\makebox(30,10){$\cdot$}}

\put(47,57){\makebox(30,10){$\cdot$}}
\put(47,55){\makebox(30,10){$\cdot$}}
\put(47,53){\makebox(30,10){$\cdot$}}
}

\put(30,70){\makebox(14,10){$N_i \! = $}}
\put(47,70){\framebox(30,10){$R_i $}}
\put(80,70){\framebox(18,10){$M_i $}}

\put(30,40){\makebox(14,10){$N_1 \! = $}}
\put(47,40){\framebox(30,10){$R_1 $}}
\put(80,40){\framebox(18,10){$M_1 $}}

\put(47,0){\framebox(30,10){$R_0$}}

\end{picture}
\label{eq:lookingglass}
\eea

\vspace*{5ex}

In Fig.~(\ref{eq:lookingglass}), RSP$_{iA}$ denotes Alice's RSP
measurement to obtain the instruction $R_i$ for Bob to prepare
$|\phi_{i}\>$.  RSP$_{iB}$ denotes Bob's conditional operation to
complete the preparation of $|\phi_{i}\>$.

The initial amount of classical communication can be created by Alice
and Bob using $cn$ uses of $U$ inefficiently, for some constant
$c$.\footnote{Sec.~\ref{sec:gbdd} shows that any nonlocal gate $U$ has
nonzero communication capacities in both directions.} 
The communication rate is 
\bea
	{ \, k n \, \Delta \mchi_U^{(1,*)} \over cn + nk} 
	\xrightarrow{k \ra \infty} \Delta \mchi_U^{(1,*)} \,.
\eea

We have not yet discussed small inaccuracies and inefficiencies in the
protocol.  The asymptotic correctness of this protocol comes from
the asymptotic reliability of its component pieces: RSP and the HSW
Thm.  However, since errors and inefficiencies accumulate over many
rounds, we need to choose the rates of increase of $n$ and $k$ slightly
more carefully.

Suppose that preparing a member of ${\cal E}^{\ot n}$ with RSP
requires $n (\mchi(\tr_{AA'} {\cal E}) + \delta^{\text{\sc rsp}}_n)$ bits
of communication and has error $\epsilon^{\text{\sc rsp}}_n$, where
$\delta^{\text{\sc rsp}}_n, \epsilon^{\text{\sc rsp}}_n \ra 0$ as
$n\ra\infty$.
Similarly, a state in $(U{\cal E})^{\ot n}$ provides 
$n(\mchi(\tr_{AA'}U {\cal E}) - \delta^{\text{\sc hsw}}_n)$ bits of
information with error $\epsilon^{\text{\sc hsw}}_n$, where again
$\delta^{\text{\sc hsw}}_n, \epsilon^{\text{\sc hsw}}_n \ra 0$ as
$n\ra\infty$.  Combining these into $\delta_n=\delta^{\text{\sc
rsp}}_n+\delta^{\text{\sc hsw}}_n$ and $\epsilon_n=\epsilon^{\text{\sc
rsp}}_n+\epsilon^{\text{\sc hsw}}_n$, we find that the communication 
rate is
\bea
	\frac{kn(\Delta\mchi_U^{(1,*)} - \delta_n)}{cn + nk}
	= \left( \Delta\mchi_U^{(1,*)} - \delta_n \right) 
	  \left( 1 - {c \over k} + O(k^{-2}) \right)
	\xrightarrow{n,k\ra\infty}
	\Delta \mchi_U^{(1,*)}
\label{eq:lglimit}
\eea
and the total error is $k\epsilon_n$.  This vanishes if one chooses
$k$ first, and then chooses $n$ such that $\epsilon_n k$ is small
($n$ thus depends on $k$).

We summarize the order of the limits.  First, choose the optimal
ensemble ${\cal E}$ to approximate $\Delta\mchi_U^{(1,*)}$.  Second,
choose $k$ large to make $c/k$ negligible (to overcome the initial
cost).  Finally choose $n$ large to make both of $k\epsilon_n$ and
$\delta_n$ vanish.
As this protocol does not require initial mutual information, it
follows that: 

{\bf Theorem 2 (2nd half):} $\lim_{t \ra \infty} \Delta
\mchi_U^{(t,\emptyset)} \geq \Delta \mchi_U^{(1,*)}$.  

Putting these together gives: 

{\bf Theorem 2:}
$	C^E_\ra = \lim_{t \ra \infty} \Delta \mchi_U^{(t,\emptyset)} 
	= \Delta \mchi_U^{(1,*)} = \sup_{\cal E} \lbm 
	\mchi(\tr_{AA'} U {\cal E}) - \mchi(\tr_{AA'} {\cal E}) \rbm \,. $

Thus initial mutual information does not increase the asymptotic
capacity, analogous to entanglement generation.  Finally, we
generalize Theorem 2 to prove strong additivity:

{\bf Theorem 2S:} The classical communication achievable by $n_i$
uses of $U_i$ is asymptotically $\sum_i n_i \, \Delta \mchi_{U_i}^{(1,*)}$.

{\bf Proof:} The argument that proves Theorem 2 (1st half) can be
applied to prove that the amount of communication generated is no more
than $\sum_i n_i \Delta \mchi_{U_i}^{(1,*)}$, which is achieved by
applying the optimal protocol for each $U_i$ separately.

\subsection{Additivity} 

We conclude this section with two observations about additivity: 

$\bullet$ We emphasize that in Theorem 2 (1st half), the Holevo bound
is applied to the output of a general protocol ${\cal P}_t$ with
possibly entangled inputs to different uses of $U$.  Thus the $1$-way
entanglement-assisted capacity for unitary bidirectional channels is
strongly additive independent of whether the Holevo information $\mchi$
is additive or superadditive.

$\bullet$ In the optimal asymptotic protocol, the $n$ copies of $U$
are applied to $n$ states each chosen from the optimal input ensemble.
Thus, entangling the inputs to different uses of $U$ does not improve
$C_{\ra,U}^E$.

\section{Discussion} 
\label{sec:diss}

Despite the many similarities between generating entanglement and
entanglement-assisted classical communication, there is an important
difference.  Communication cannot be stored and be used later.  In
particular, Alice needs to work backwards in our optimal
entanglement-assisted communication protocol, so that the classical
messages need to be known at the beginning of the protocol to share
the initial cost.
In contrast, entanglement can be stored.  The optimal entanglement
generation protocol can be stopped and resumed at arbitrary times.

We can generalize the first half of Theorems 1 and 2 to any other
quantity which is monotonic under the given resources, as long as a
sufficiently general input (e.g. state or ensemble) is allowed for the
$1$-shot capacity.  In particular, the input should possess all the
properties the output may possess.
If in addition the quantity is weakly additive or subadditive on the
optimal input and weakly additive or superadditive on the optimal
output, repeating the optimal $1$-shot protocol allows the upper bound
to be attained asymptotically, and additivity holds.

We end this section with a discussion on the parallel versus sequential
applications of bidirectional channels in a protocol.  
Note that there is no such distinction for unidirectional channels (in
the absence of back channels), as the output state of a given
application of the channel is with the receiver and can never be used
as an input for later uses.
For bidirectional channels, there are sequential schemes that cannot 
be made parallel.  
For example, the protocol for entanglement-assisted $1$-way classical
communication in \sec{lglass} cannot be made parallel.
Sequential schemes are always at least as powerful as parallel ones.  
The opposite is true in the asymptotic regime, in which case any
capacity of $U^{\ot n}$ (i.e. one must apply $n$ copies of $U$ in
parallel) is equal to $n$ times the capacity of $U$.  The proof is
simple -- let ${\cal P}_t$ be any protocol that uses $U$ sequentially.
A particular $t$-use protocol for $U^{\ot n}$ is to run $n$ copies of
${\cal P}_t$ in parallel.  Thus the $t$-shot capacity of $U^{\ot n}$
is no worse than $n$ times that of $U$, and equality holds.

\section{Other general bounds}
\label{sec:gbdd}

We have proved a few simple general bounds: 
$E_U^{(*,cc)} \leq$ $\log$ Sch$(U) \leq 2 \log d$ and 
$C^E_{\ra,U} \,, C^E_{\la,U} \leq 2 \log d$. 
We now derive other general bounds that hold for all $U$.  
We focus on the entropy of entanglement $E_e$, and on the two
capacities $E_{e,U}^{(1,\emptyset)}$ and $E_{e,U}^{(\emptyset)}$ ($=
E_{e,U}^{(1,*)})$ since the latter is equal to many entanglement
capacities of our interest (see \sec{esum}).

\medskip

{\bf Bound 1:} 
$U$ is nonlocal $~\Leftrightarrow~ E_{e,U}^{(1,\emptyset)} > 0
~\Leftrightarrow~ C_{\ra,U} > 0 ~\Leftrightarrow~ C_{\la,U} > 0$.

{\bf Proof:} The first equivalence follows from Corollary 3.5 in
\sec{asymres}.  Let $E_0 > 0$ be the amount of entanglement created by
applying $U$ to $|\Phi_d\>_{AA'}|\Phi_d\>_{BB'}$.

Alice can send a noisy bit to Bob with the following $t$-use protocol.
Bob inputs $|\Phi_d\>_{BB'}^{\otimes t}$ to all $t$ uses of $U$. 
To send ``$0$'' Alice inputs $|\Phi_d\>_{AA'}^{\otimes t}$ to share $t
E_0$ ebit with Bob.  To send ``$1$'', Alice inputs $|0\>_A$ to
the first use of $U$, takes the output and uses it as the input to the
second use, and so on, so that their final entanglement is no more
than $\log d$.
Thus different messages from Alice result in very different amount of 
entanglement at the end of the protocol.
Using Fannes' inequality~\cite{Fannes73,Nielsen99c}, 
$t E_0 - \log d \leq t \log d \; \tr|\rho_{0}-\rho_{1}| + {1 \over
e}$ where $\rho_{i}$ is the reduced density matrices of Bob when Alice
sends $i$.
For any $\epsilon > 0$, $\exists \, t$ such that $E_0 - \epsilon \leq
\log d \; \tr|\rho_{0}-\rho_{1}|$ and Bob can distinguish $\rho_0$
from $\rho_1$ with nonzero advantage.  It means that the $t$-use
protocol then simulates a noisy classical channel with nonzero
capacity and $C_{\ra,U} > 0$.  Obviously $C_{\ra,U} > 0$ implies $U$
is nonlocal.  Similarly $C_{\ra,U} > 0 \; \Leftrightarrow \; U$ is
nonlocal. \qed

\medskip

{\bf Bound 2:} $E_U^{(\emptyset)} \geq C_{+,U}$.  


{\bf Proof:} 
Suppose a $t$-use protocol ${\cal P}_t$ transmits $n_a$ bits from
Alice to Bob and $n_b$ bits from Bob to Alice with fidelity
$1-\epsilon$.  
Recall from \sec{def} that ${\cal P}_t$ can be assumed unitary with
the ancillas starting in the state $|0r\>_{A'} |0r\>_{B'}$ where $r$
is a share random variable.
Let $|x\>_A |y\>_B$ carry the messages to be communicated, where $x$
and $y$ are $n_a$- and $n_b$-bit strings.
Then, by definition (\eq{C_U-def}), the state change is given by: 
\bea
      & & |x\>_A |y\>_B |0r\>_{A'} |0r\>_{B'} 
	~ \stackrel{{\cal P}_t}{\longrightarrow} ~ |\eta_{xy}\>
\non
\\
	& {\rm s.t.} & \forall_{x,y} ~
	F \, ( \,|y\>_A |x\>_B \,, \tr_{\!A'\!B'} 
	|\eta_{xy}\>\<\eta_{xy}|_{A\!B\!A'\!B'} \,) 
	= 1 - \epsilon_{xy} > 1-\epsilon 
\,.
\eea
By Uhlmann's Theorem \cite{Uhlmann76}, there are normalized states  
$|c_{xy}\>_{A'\!B'}$ and $|e_{xy}\>_{A\!B\!A'\!B'}$ such that
\bea
	|\eta_{xy}\> = \sqrt{1-\epsilon_{xy}} \;  
		  |y\>_A |x\>_B |c_{xy}\>_{A'\!B'}
		+ \sqrt{\epsilon_{xy}} \;  |e_{xy}\>_{A\!B\!A'\!B'}
\,, 
\eea
and ${\rm tr}_{A'\!B'} |e_{xy}\>\<e_{xy}|_{A\!B\!A'\!B'}$ has support
orthogonal to the span of $|y\>_A |x\>_B$.

%
%

To prove $E_U^{(\emptyset)} \geq C_{+,U}$, we simply change the inputs to
the protocol so that it creates entanglement.
Alice's input system $A$ is now in a maximally entangled state with
another ancilla $A''$, each with $2^{n_a}$ dimensions, and similarly
for Bob.
Thus the input state is given by 
\bea
	2^{-(n_a+n_b)/2} \sum_{xy}
	|x\>_A |x\>_{A''} |y\>_B |y\>_{B''} |0r\>_{A'} |0r\>_{B'}
\,,
\eea
where $x$ and $y$ are summed over their possible values. 
The output is given by 
\bea
	|\eta^{(\epsilon)}\> & = & 2^{-(n_a+n_b)/2} \sum_{xy} |\eta_{xy}\>
\non
\\
	& = & 2^{-(n_a+n_b)/2} \sum_{xy} 
	\lbL \sqrt{1-\epsilon_{xy}} \;  |y\>_A |x\>_B |c_{xy}\>_{A'\!B'}
		+ \sqrt{\epsilon_{xy}} \;  |e_{xy}\>_{A\!B\!A'\!B'} \rbL 
	|x\>_{A''} |y\>_{B''} 
\label{eq:etaep}
\,.
\eea
To calculate $E(|\eta^{(\epsilon)}\>)$, we first calculate
$E(|\eta^{(0)}\>)$ for
\bea
	|\eta^{(0)}\> = 2^{-(n_a+n_b)/2} \sum_{xy}
	|y\>_A |x\>_{A''} |x\>_B |y\>_{B''} |c_{xy}\>_{A'B'}
\label{eq:eta0}
\,.  
\eea
$E(|\eta^{(0)}\>)$ is simply the entropy of Alice's reduced density
matrix, which can be found by the ``Joint Entropy Theorem''
(Eq.~(1.58) in \cite{Nielsen00}).
\bea 
	S \lpm \tr_{BB'B''} |\eta^{(0)}\>\<\eta^{(0)}| \rpm & = & 
	S \lpL 2^{-(n_a+n_b)} \sum_{x y} 
	|y\>\<y|_A |x\>\<x|_{A''} \otimes \tr_{B'} 
	|c_{xy}\>\<c_{xy}| \rpL
\non
\\
	& = & n_a + n_b + 2^{-(n_a+n_b)} 
	\sum_{xy} S \lpL \tr_{B'} |c_{xy}\>\<c_{xy}| \rpL 
	\geq n_a + n_b
\,.
\label{eq:newpdd}
\eea
We now relate $E(|\eta^{(\epsilon)}\>)$ to $E(|\eta^{(0)}\>)$.
The inner product $\<\eta^{(0)}|\eta^{(\epsilon)}\>$ can be bounded: 
\bea
	\<\eta^{(0)}|\eta^{(\epsilon)}\> 
	& = & 2^{-(n_a+n_b)} \sum_{xy}
	\<y|_A \<x|_B \<c_{xy}|_{A'\!B'} ~
	\lbm \sqrt{1-\epsilon_{xy}} \;  |y\>_A |x\>_B |c_{xy}\>_{A'\!B'}
		+ \sqrt{\epsilon_{xy}} \;  |e_{xy}\>_{A\!B\!A'\!B'} \rbm
\non  
\\
	& = & 2^{-(n_a+n_b)} \sum_{xy} \sqrt{1-\epsilon_{xy}} 
\non
\\
	& \geq & \sqrt{1-\epsilon} 
\label{eq:epsilonbdd}
\eea
where we have used the the orthogonality of $\{ |x\>_{A''} |y\>_{B''}
\}_{xy}$ to obtain the first line, and the orthogonality of 
$|y\>_A |x\>_B |c_{xy}\>_{A'\!B'}$ and $|e_{xy}\>_{A\!B\!A'\!B'}$ 
to obtain the second line.  
The trace distance $D$ between $|\eta^{(0)}\>$ and
$|\eta^{(\epsilon)}\>$ is defined to be ${1 \over 2} {\rm tr}
|~|\eta^{(0)}\>\<\eta^{(0)}|-|\eta^{(\epsilon)}\>\<\eta^{(\epsilon)}|~|$,
and $D(|\eta^{(0)}\>,|\eta^{(\epsilon)}\>) =
\sqrt{1-\<\eta^{(0)}|\eta^{(\epsilon)}\>^2} \leq \sqrt{\epsilon}$ (see
Sec.~9.2.3 of \cite{Nielsen00}).
Using Fannes' inequality \cite{Fannes73} or the continuity of the
entropy of entanglement~\cite{Nielsen99c}, 
\bea
	|~E(|\eta^{(0)}\>) - E(|\eta^{(\epsilon)}\>)~| & \leq &  
	\log \ss \lpm 
	{\rm Sch}(|\eta^{(0)}\>) + {\rm Sch}(|\eta^{(\epsilon)}\>) \rpm 
	~2 D(|\eta^{(0)}\>,|\eta^{(\epsilon)}\>) + 1/e 
\\
	& \leq & (2 t \log d + n_a + n_b) ~2 \sqrt{\epsilon} + 1/e 
\,.
\label{eq:badfannes}
\eea
We explain how \eq{badfannes} is obtained.  First, ${\rm
Sch}(|\eta^{(\epsilon)}\>) \leq d^{2t}$ because the protocol has a
product initial state and Sch$(U|\psi\>) \leq$ Sch$(U)$
Sch$(|\psi\>)$~\cite{Nielsen98}.
We now bound ${\rm Sch}(|\eta^{(0)}\>)$.  
A Schmidt decomposition of $|\eta^{(0)}\>$ can be obtained by Schmidt
decomposing each $|c_{xy}\>$ in \eq{eta0} so that ${\rm
Sch}(|\eta^{(0)}\>) = \sum_{xy} {\rm Sch}(|c_{xy}\>)$.
For each $xy$, ${\rm Sch}(|c_{xy}\>) \leq {\rm Sch}(|\eta_{xy}\>)$
since $|c_{xy}\>$ can be obtained with nonzero probability by locally
measuring $|\eta_{xy}\>$.
Each $|\eta_{xy}\>$ is obtained from a product initial state
after $t$ applications of $U$, and ${\rm Sch}(|\eta_{xy}\>) \leq
d^{2t}$.
Altogether, ${\rm Sch}(|\eta^{(0)}\>) \leq d^{2t} \, 2^{n_a +n_b}$ 
and ${\rm Sch}(|\eta^{(0)}\>) + {\rm Sch}(|\eta^{(\epsilon)}\>) 
\approx d^{2t} \, 2^{n_a +n_b}$.

{From} \eq{newpdd} and \eq{badfannes} we can lower bound the
entanglement generated per use of $U$:
\[
{1 \over t} E(|\eta^{(\epsilon)}\>) \; \geq \; {1 \over t} (n_a + n_b)
- 2 \sqrt{\epsilon} \; \lbm 2 \, \log d + {1 \over t}(n_a +n_b) \rbm -
{1 \over et}
\] 
As $t$ increases, $\epsilon$ can be made arbitrarily small and ${1
\over t} (n_a + n_b) \rightarrow C_{+,U}$.  Furthermore, $2 \log d +
(n_a +n_b)/t \ra 2 \log d + C_{+,U} \leq 4 \log d$ is well bounded.
The above equation then implies $E_U^{(\emptyset)} \geq C_{+,U}$. \qed

{\bf Remark:} In the proof above, it is crucial to bound
Sch$(|\eta^{(\epsilon)}\>)$ and Sch$(|\eta^{(0)}\>)$ as functions of
$t$, and our bound is based on having a product initial state.
Furthermore, the two limits $t \ra \infty$ and $\epsilon \ra 0$ are
dependent.  Thus one cannot assume $\epsilon \ra 0$ as a separate
premise in the above proof and extra care is needed in how the limits
are taken.

After this paper was first posted, Berry and Sanders \cite{Berry02a}
proved that if the capacity is achievable by an exact protocol
(i.e. $\epsilon = 0$), then $C_{+,U}^E \leq E_U^{(\emptyset)} +
E_{U^\dagger}^{(\emptyset)}$.

To adapt the proof of bound 2 for $C_{+,U}^E \leq E_U^{(\emptyset)} +
E_{U^\dagger}^{(\emptyset)}$ in the general case when $\epsilon > 0$
will require an explicit bound on Sch$(|\eta^{(\epsilon)}\>)$ and
Sch$(|\eta^{(0)}\>)$ and knowledge of how various inaccuracies vanish
asymptotically, so as to specify how various dependent limits should
be taken.  So far, we do not see how this can be done.

In the following, we prove a weaker bound $C_{\ra,U}^E \leq
E_U^{(\emptyset)} + E_{U^\dagger}^{(\emptyset)}$ for $\epsilon > 0$,
by adapting the proof of bound 2 and an idea from \cite{Berry02a}, as
well as using details on the optimal protocol for achieving
$C_{\ra,U}^E$ and an improved method for RSP of bipartite pure
entangled state that uses less entanglement than the method in
Ref.~\cite{peterrsp}.

Before we present the proof, we give an interpretation of
$E_{U^\dagger}^{(\emptyset)} = E_{U^\dagger}^{(1,*)}$ as the {\em
entanglement destroying capacity} of $U$
\bea
	\sup_{|\psi\>_{AA'BB'}} \lbm E(|\psi\>_{AA'BB'}) -
	E(U_{AB} |\psi\>_{AA'BB'})  \rbm \,, 
\eea
since $U^\dagger$ creates as much entanglement on the input $|\psi\>$
as $U$ can destroy on $U^\dagger|\psi\>$.  Note that to disentangle a
state {\em unitarily} is a nonlocal task.  We now turn to our proof. 

\medskip

{\bf Bound 3:} $C_{\ra,U}^E \leq E_U^{(\emptyset)} +
E_{U^\dagger}^{(\emptyset)}$.

{\bf Proof:} We omit details already given in the proof of bound 2.
Let ${\cal P}_t$ be a unitary protocol transmitting $n_a$ bits from
Alice to Bob with fidelity $1-\epsilon$.
The ancillas are initially in the maximally entangled state
$|\Phi_M\>_{A'B'}$ where $\log M$ is the amount of initial
entanglement required to assist the communication.
Let $|x\>_A$ carry the $n_a$-bit message of Alice.
The state change is given by: 
\bea
      & & |x\>_A |0\>_B |\Phi_M\>_{A'B'} 
	~ \stackrel{{\cal P}_t}{\longrightarrow} ~ |\eta_{x}\>
\\
      & & |\eta_{x}\> = \sqrt{1-\epsilon_x} \;  |0\>_A |x\>_B |c_{x}\>_{A'\!B'}
			+ \sqrt{\epsilon_x} \;  |e_{x}\>_{A\!B\!A'\!B'}
\label{eq:etax}
\, 
\eea
where ${\rm tr}_{A'\!B'} |e_{x}\>\<e_x|_{A\!B\!A'\!B'}$ has support 
orthogonal to $|0\>_A |x\>_B$.

In the entanglement generation protocol, Alice inputs half of
$|\Phi_{2^{n_{\!a}}}\>_{AA''}$ while Bob still inputs $|0\>_B$.
The input and output states are given by 
\bea
	& & 2^{-n_a/2} \sum_{x}
	|x\>_A |x\>_{A''} |0\>_B |\Phi_M\>_{A'B'} ~~~~\mbox{and} 
\\
	|\eta^{(\epsilon)} \> & = & 2^{-n_a/2} \sum_{x} 
	\lbL \sqrt{1-\epsilon_x} \;  |0\>_A |x\>_B |c_{x}\>_{A'\!B'}
		+ \sqrt{\epsilon_x} \;  |e_{x}\>_{A\!B\!A'\!B'} \rbL 
	|x\>_{A''} 
\,.
\eea
For $|\eta^{(0)}\> = 2^{-n_a/2} \sum_{x} |0\>_A |x\>_{A''} |x\>_B
|c_{x}\>_{A'B'}$,
\bea 
	E(|\eta^{(0)}\>) & = & S \lpm \tr_{BB'} |\eta^{(0)}\>\<\eta^{(0)}| 
	\rpm =  
	S \lpL 2^{-n_a} \sum_{x} 
	|0\>\<0|_A |x\>\<x|_{A''} \otimes \tr_{B'} 
	|c_{x}\>\<c_{x}| \rpL
\non
\\
	& = & n_a + 2^{-n_a} 
	\sum_{x} S \lpL \tr_{B'} |c_{x}\>\<c_{x}| \rpL
\non
\\
	& = & n_a + 2^{-n_a} 
	\sum_{x} E(|c_{x}\>)
\,.
\non
\eea
Applying the definition of the entanglement destroying capacity to 
\eq{etax}, 
\bea
	\forall_x 
	~E(|\eta_{x}\>) \geq \log M - t E_{U^\dagger}^{(1,*)} = 
	\log M - t E_{U^\dagger}^{(\emptyset)}
\,.
\eea
Since $\<0|_A \<x|_B \<c_x|_{A'\!B'} |\eta_x\>_{A\!A'\!B\!B'} \geq
\sqrt{1-\epsilon}$ and ${\rm Sch}(|c_x\>) \leq {\rm Sch}(|\eta_x\>)
\leq M d^{2t}$, Fannes' inequality implies
\bea
	& \forall_x & 
	|~E(|c_x\>) - E(|\eta_{x}\>)~| \leq 
	4 \sqrt{\epsilon} \lbm \log M + 2t \log d \rbm + 1/e
\,.
\eea
Hence, 
\bea 
	E(|\eta^{(0)}\>) \geq n_a + 
	\log M - t E_{U^\dagger}^{(\emptyset)} 
	- 4 \sqrt{\epsilon} \lbm \log M + 2t \log d \rbm - 1/e
\,.
\eea

Using $\<\eta^{(0)}|\eta^{(\epsilon)}\> \geq \sqrt{1-\epsilon}$, and
that ${\rm Sch}(|\eta^{(\epsilon)}\>) \leq d^{2t} M$, ${\rm
Sch}(|\eta^{(0)}\>) \leq \sum_x {\rm Sch}(|c_x\>) \leq 2^{n_a} 
d^{2t} M$, Fannes' inequality implies
\bea
	|~E(|\eta^{(0)}\>) - E(|\eta^{(\epsilon)}\>)~| & \leq & 
	2 \sqrt{\epsilon} 
	\lbm \log M + 2t \log d + n_a \rbm + 1/e
\,.
\eea
Thus
\bea
	E(|\eta^{(\epsilon\>)}) \geq n_a + \log M 
	- t E_{U^\dagger}^{(\emptyset)} 
	- 2 \sqrt{\epsilon} \lbm 3 \log M + 6t \log d + n_a \rbm - 2/e
\,, 
\eea
and 
\bea
	E_U^{(\emptyset)} \geq 
	{1 \over t} \lbm E(|\eta^{(\epsilon)}\>) - \log M \rbm 
	\geq {1 \over t} n_a - E_{U^\dagger}^{(\emptyset)} 
	- 2 \sqrt{\epsilon} \lbm {3 \over t} \log M + 6 \log d 
	+ {n_a \over t} \rbm + {2 \over et}
\,.
\label{eq:hell}
\eea

In particular, consider the entanglement-assisted communication
protocol in \sec{lglass}.  For any $\delta_{{\cal E}} > 0$, $\exists
{\cal E}$ consisting of $M_0$-dimensional bipartite states with
$\mchi(U {\cal E})- \mchi({\cal E}) \geq \Delta \mchi_U^{(1,*)} -
\delta_{{\cal E}} = C_{\ra,U}^E - \delta_{{\cal E}}$.  Following
\eq{lglimit}, $t = cn+nk$ for some constant $c$ and the rate is
${n_a \over t} \geq C_{\ra,U}^E - \delta_{{\cal E}} - \delta_n - 
{c \over k} \ss C_{\ra,U}^E$, where $\delta_n
\ra 0$ as $n \ra \infty$.  The total error is $\epsilon = k
\epsilon_n$, where $\lim_{n \ra \infty} \epsilon_n = 0$.
The RSP method in \cite{peterrsp} can be improved \cite{Harrowrsp} to
prepare $n$ states from an ensemble ${\cal E}$ with
$n \mchi(\tr_{AA'}{\cal E}) + o(n)$ cbits and 
$n S(\bar{\rho}_B) + o(n) \leq n \log M_0$ ebits where $\bar{\rho}_B$
is the average reduced density matrix of the ensemble as seen from
Bob, so that $M \leq M_0^t$.
Putting all these parameters into \eq{hell}, 
\bea
	E_U^{(\emptyset)} 
	\geq C_{\ra,U}^E - \lbm 
	\delta_{{\cal E}} + \delta_n + {c \over k} \ss C_{\ra,U}^E \rbm 
	- E_{U^\dagger}^{(\emptyset)} 
	- 2 \sqrt{\epsilon} \lbm 3 \log M_0 + 6 \log d + {n_a \over t} 
	\rbm + {2 \over e n(c+k)}
\label{eq:hellhell}
\eea
For any $\delta > 0$, choose \\
1. ${\cal E}$ such that $\delta_{{\cal E}} < \delta/5$, \\ 
2. $k$ such that ${c \over k} \ss C_{\ra,U}^E < \delta/5$ and 
$2/e(c+k) \leq \delta/5$ so that $2/en(c+k) \leq \delta/5$, \\
3. $n$ such that $\delta_n < \delta/5$ and $\epsilon = k \epsilon_n$ small 
enough for $2 \sqrt{\epsilon} \lbm 3 \log M_0 + 6 \log d + {n_a \over t}
\rbm < \delta/5$. 
\qed 

\medskip

To summarize, for all $U$, we have \\
$\bullet$ $C_{\ra,U}, \, C_{\la,U} \leq C_{+,U} \leq E_{c,U}^{(1,*)} 
\leq$ Sch$(U) \leq 2 \log d$. \\
$\bullet$ 
$C_{+,U} \leq C_{\ra,U} \! + \! C_{\la,U}$,  
$C_{+,U} \leq E_U^{(\emptyset)}$, 
$C_{+,U} \leq C_{+,U}^{E}$. 
\\
$\bullet$
$C_{\la,U} \leq C^{E}_{\la,U} \leq \min( \, 2 \log d \,, \; 
E_U^{(\emptyset)} \! + \! E_{U^\dagger}^{(\emptyset)})$,
$C_{\ra,U} \leq C^{E}_{\ra,U} \leq \min( \, 2 \log d \,, \; 
E_U^{(\emptyset)} \! + \! E_{U^\dagger}^{(\emptyset)})$, 
\\
$~~~\, C^{E}_{+,U} \leq C^E_{\ra,U} + C^E_{\la,U} \leq 
\min(4 \log d, 2 \ss (E_U^{(\emptyset)} + E_{U^\dagger}^{(\emptyset)}))$.
\\
$\bullet$
$C_{\lra,U} \leq \min (C_{\ra,U}, C_{\la,U})$, $C_{+,U} 
\geq 2 \ss C_{\lra,U}$.

\medskip

We now return briefly to Hamiltonian capacities.  Recall from
\sec{structure} that any Hamiltonian capacity can be expressed in
terms of the corresponding gate capacity,
$G_H = \lim_{s \ra 0} {1 \over s} \; G_{U = e^{-iHs}}$.
The finiteness of $G_H$ is not immediate from the above definition.
Even though $G_{U = e^{-iHs}} \ra 0$ when $s \ra 0$ due to continuity, 
it is not guaranteed that $G_{U = e^{-iHs}} \leq O(s)$.  
One may argue that physically, the rate should be finite, but the 
availability of unlimited local resources complicates the argument.  
We now provide a proof of the finiteness of the Hamiltonian
capacities~\cite{Guifre02}.

{\bf Bound 4:} Hamiltonian capacities are finite.  

{\bf Proof:} Recall that the entanglement capacity of a gate $U$ is no
more than the average amount of entanglement required to simulate $U$
given free classical communication.  Reference \cite{Cirac00}
describes a method to simulate $U = e^{-i \alpha \sigma_z \ot
\sigma_z}$ with $O(\alpha)$ entanglement, which implies that $E_{H =
\sigma_z \ot \sigma_z}$ is finite.  Reference \cite{Bennett01}
describes a method to simulate any other Hamiltonian $H'$ in $d \times
d$ using the Hamiltonian $H = \sigma_z \ot \sigma_z$ with $O(d^4)$
overhead, so that $\forall_{H'}$ $E_{H'} \leq O(d^4) E_{H=\sigma_z \ot
\sigma_z}$ is also finite.  


The finiteness of other capacities now follows.
{From} the summary, $C_{+,H'}^{E} \leq 2(E_{H'} + E_{-H'})$ is an
upper bound to all other classical capacities of $H'$. \qed

\section{Bidirectional channels on $d_1 \times d_2$}

We have assumed $U$ acting on a $d \times d$ bipartite system.  We
note here that all the results discussed hold for a nonlocal gate (or
Hamiltonian) acting on a $d_1 \times d_2$ system (without loss of
generality, $d_1 \leq d_2$).  The interested reader can easily verify
that all the arguments hold in this case, because the fact $d_1 = d_2$
is {\em never} used in the proofs.  We also note a subtle observation,
that the $d_1 \times d_2$ case is not described by embedding the
operation in a $d_2 \times d_2$ system by taking the direct sum
with a $d_2-d_1$ dimensional identity matrix acting on the side of 
lower dimension.

\section{Open questions and examples}
\label{sec:evilgates}

We have found expressions for the entanglement capacity and the
entanglement-assisted classical capacity of unitary bidirectional
channels, defined classical capacities for them, and provided general
bounds for the capacities.  We conclude first with a list of open
questions, followed by examples to illustrate our results and our open
questions.

\subsection{Open questions}  

$\bullet$ How large do the ancillas $A'B'$ need to be in the optimal 
input for entanglement generation?  How large do $A'B'$ 
need to be, and how many states are needed in the optimal ensemble for
entanglement-assisted classical communication?
These are important for numerical studies of the capacities.  

$\bullet$ Will infinite dimensional ancillas improve the entanglement
capacity and the entanglement-assisted $1$-way classical capacity?
Will an ensemble with an infinite number of members improve the
latter?

$\bullet$ How do the forward and backward rates trade off with
each other (in either the unassisted or assisted case)?

$\bullet$ Are forward and backward classical capacities always equal
(in either the unassisted or assisted case)?

$\bullet$ Is there a gate $U$ with $C_{+,U} < E_{e,U}^{(\emptyset)}$ a
strict inequality?

$\bullet$ Is $E_{e,U}^{(1,*)} = E_{e,U^\dagger}^{(1,*)}$ for all $U$?
Both quantities relate to how entangling a nonlocal gate is.  However,
we can only prove the equality when $U = U^T$, by using the fact
$E_{e,U}^{(1,*)} = E_{e,U^*}^{(1,*)}$ ($U^*$ is the complex conjugate
of $U$).
This generalizes the proof in Ref.~\cite{Berry02a} for $2$-qubit gates
since $U = U^T$ for all $2$-qubit gates in their normal
form~\cite{Kraus00}.  
Numerical work suggests that the equality does not hold for some $U$ 
in higher dimensions~\cite{CDS}.

$\bullet$ When can a gate be simulated efficiently, i.e., by an amount
of some resource equal to the capacity?

$\bullet$ How do auxiliary resources of quantities linear in the
number of uses affect the capacity?

\subsection{Examples} 

{\bf Example 1:} Let $U = \mbox{\sc cnot}$.  It can be simulated using
$1$ ebit and $1$ bit of classical communication in each
direction~\cite{Gottesman98}.  Thus
$E_{\mbox{\sc cnot}}^{(\emptyset)} \leq 1$, 
$C^{E}_{\ra, \mbox{\sc cnot}} \leq 1$, 
$C^{E}_{\la, \mbox{\sc cnot}} \leq 1$, and  
bound 2 further implies $C_{+,\mbox{\sc cnot}} \leq 1$.
These are all achievable with obvious methods, without the need of
entanglement assistance in $C^{E}_{\ra, \mbox{\sc cnot}}$ and
$C^{E}_{\la, \mbox{\sc cnot}}$ and without the need of initial
entanglement in $E_{\mbox{\sc cnot}}^{(\emptyset)}$.
Therefore, $C_{\ra, \mbox{\sc cnot}} = C_{\la, \mbox{\sc cnot}} =
C_{+, \mbox{\sc cnot}} = C^{E}_{\ra, \mbox{\sc cnot}} = C^{E}_{\la,
\mbox{\sc cnot}} = E_{\mbox{\sc cnot}}^{(\emptyset)} = E_{\mbox{\sc
cnot}}^{(1,\emptyset)} = 1$.
The rate pairs in the triangle with vertices $(0,0)$, $(0,1)$, $(1,0)$
are achievable without entanglement assistance, and convexity implies
no other pair is achievable.
We also have $C^{E}_{+,\mbox{\sc cnot}} \geq 2$ due to the following
protocol.
Starting with the EPR state $|\Phi_2\>_{AB}$, Alice applies $\sigma_x^a$
and Bob applies $\sigma_z^b$ if their respective input bits are $a$
and $b$.  
The {\sc cnot} is then applied, converting the state to ${(-1)^{ab}
\over \sqrt{2}} \lpm |0\>+(-1)^b|1\> \rpm |a\>$.
Thus rate pairs in the square with vertices $(0,0)$, $(0,1)$, $(1,0)$,
$(1,1)$ are all achievable with entanglement assistance, and by
monotonicity, no other rate pair is achievable.

\medskip

{\bf Example 2:} 
For $U = \mbox{\sc swap}$, we have the general upper bounds 
$E_{\mbox{\sc swap}}^{(\emptyset)} \leq 2$, 
$C^{E}_{\ra, \mbox{\sc swap}} \leq 2$,  
$C^{E}_{\la, \mbox{\sc swap}} \leq 2$, and bound 2 implies  
$C_{+, \mbox{\sc swap}} \leq 2$.  
These are all achievable as follows.  
$E_{\mbox{\sc swap}}^{(\emptyset)} = E_{\mbox{\sc
swap}}^{(1,\emptyset)}$ is achieved on the input $|\Phi_2\>_{A\!A'}
|\Phi_2\>_{B\!B'}$.
To achieve the forward assisted classical capacity, Alice and Bob
start with the state $|\Phi_2\>_{A\!B''} |\Phi_2\>_{B\!B'}$ and Alice
applies $\sigma_{xA}^{a_1} \sigma_{zA}^{a_2}$ when her $2$-bit message
is $a_1 a_2$.  Then {\sc swap} is applied.  In other words, superdense
coding~\cite{Bennett92} is performed, consuming an existing EPR pair
on $A\!B''$, while a new EPR pair is created on $A\!B'$
simultaneously.  Thus the unassisted and assisted one-way classical
capacities are both $2$.  $C_{+, \mbox{\sc swap}} = 2$ is achieved in
the obvious way.
Superdense coding in both the forward and backward directions implies
$C^{E}_{+, \mbox{\sc swap}} \geq 4$, and by the monotonicity of the
achievable region of assisted rate pairs, $C^{E}_{+, \mbox{\sc swap}}
= 4$.
Therefore, any rate pair inside the triangle with vertices
$(0,0),(0,2),(2,0)$ can be achieved without entanglement assistance,
and any rate pair inside the square with vertices
$(0,0),(0,2),(2,0),(2,2)$ can be achieved with
entanglement assistance.

The {\sc cnot} and {\sc swap} are very simple.  
We now turn to more intriguing examples. 

\medskip

{\bf Example 3:} 
The gate {\sc j} acts as 
\[
	|00\> \leftrightarrow {1 \over \sqrt{2}} (|00\>+|11\>)	\,,
\hspace*{5ex}
	|11\> \leftrightarrow {1 \over \sqrt{2}} (|00\>-|11\>)	\,,
\hspace*{5ex}
	|01\> \leftrightarrow |01\> \,,
\hspace*{5ex}
	|10\> \leftrightarrow |10\> 
\]
where the first and second registers are $A$ and $B$ (same throughout 
the examples).  
Without ancillas, {\sc j} creates $1$ ebit but seems to create less
than $1$ cbit in $1$-shot, but \cite{Bernstein02} presents a product
$2$-qubit input that communicates $1$ cbit from Alice to Bob.

Numerical optimization of the generated entanglement with
$2$-dimensional $A'$ and $B'$ in \eq{e1} is $1.83186$ ebit, and the
optimal input has $0.055338$ ebit.  As a comparison, only $1.8113$
ebit is generated by inputting $|\Phi_2\>_{A\!A'} |\Phi_2\>_{B\!B'}$.
%
%

Starting from $|\Phi_2\>_{AB} = {1 \over \sqrt{2}} ( |00\>+|11\> )$, Alice
and Bob can communicate one bit to each other, by applying
$\sigma_z^a$ and $\sigma_z^b$ if their respective messages are $a$ and
$b$.  The {\sc j} gate further converts the state to $|x\>_A |x\>_B$
where $x = a+b \mod 2$, from which they learn each other's input.

We suspect $C_{+,\mbox{\sc j}} < E_{\mbox{\sc j}}^{(\emptyset)}$.
For instance, the best total rate we found requires creating $1$ ebit
with $1$ use of {\sc j} followed by assisted two-way communication in
the second use of {\sc j}.  Asymptotically, $2.83186$ uses of {\sc j}
can create at least $1.83186 \times 2$ bits of communication, so that
$C_{+,\mbox{\sc j}} \geq 1.2938$, which is much less than $1.83186$.

\medskip

{\bf Example 4:} 
Denote the ``cyclic permutation'' gate by {\sc cp}.  It acts as: 
\bea
	|x\> |y\> ~ \ra & \hspace*{-3.2ex} 
	|x\> |y\!-\!1\> & {\rm ~if~} y \neq 0
\non
\\
	|x\> |y\> ~ \ra & |x\!-\!1\> |y\!-\!1\> & {\rm ~if~} y = 0
\non
\eea
where $\pm$ is modulo $d$.  
Sch$(${\sc cp}$) = 2$ for all $d$, thus $E_{\mbox{\sc
cp}}^{(\emptyset)} \leq 1$.  This is achievable on the input ${1
\over \sqrt{2}} |0\>_A (|0\>+|1\>)_B$.
$C_{\la,\mbox{\sc cp}} \geq 1$ since Bob can send $1$ bit $b \in
\{0,1\}$ to Alice with the input $|0\>_A |b\>_B$.
Thus $\forall_d$ $E_{\mbox{\sc cp}}^{(\emptyset)} = E_{\mbox{\sc
cp}}^{(1,\emptyset)} = C_{\la,\mbox{\sc cp}} = C_{+,\mbox{\sc
cp}} = 1$.

For $d = 3$, we have also studied forward communication without
ancillas.  It is impossible to transmit $1$ bit from Alice to Bob by
one use of $U$, but it is possible asymptotically, so that
$C_{\ra,\mbox{\sc cp}} = 1$.

\medskip

{\bf Example 5:} 
Define the gate {\sc a\!e} on $d \times d$ by: 
\bea
\non 
	& \forall ~x \hspace*{17ex} &
	|x0\> \lra |xx\> \,,  
\\
\non
	& \forall ~0 \neq y ~~{\rm and}~~ y \neq x & 
	|xy\> \lra |xy\> \,.   
\eea

Since Sch$(U) = d$, $E_{\mbox{{\sc a\!e}}}^{(\emptyset)} \leq \log d$,
and this is achievable on the input $\sum_x |x\>_A |0\>_B$, and 
$E_{\mbox{{\sc a\!e}}}^{(\emptyset)} = E_{\mbox{{\sc
a\!e}}}^{(1,\emptyset)} = \log d$.
$C_{\ra,\mbox{{\sc a\!e}}} = \log d$ is achievable in the obvious
manner.  Thus $C_{+,\mbox{{\sc a\!e}}} = \log d$.
We can prove that one use of {\sc a\!e} can communicate strictly less 
than $\log d$ cbit from Bob to Alice starting from product states but 
allowing ancillas. 
However, we suspect $C_{\la,\mbox{{\sc a\!e}}} < C_{\ra,\mbox{{\sc
a\!e}}} = \log d$.

\medskip

{\bf Example 6:} Since the initial submission of this paper, Childs,
Leung, Verstraete, and Vidal~\cite{Childs02e1} have analytically
proved that the asymptotic entanglement capacity of any Hamiltonian
locally equivalent to $\alpha \, \sigma_x \ot \sigma_x + \beta \,
\sigma_y \ot \sigma_y$ can be achieved without ancillas, and the
capacity is $\approx 1.9123 \, (\alpha + \beta)$ following
Ref.~\cite{Dur00}.

\section{Acknowledgements}

We thank M. Leifer, L. Henderson, and N. Linden for discussions and
for kindly sharing their results on entanglement capacity prior to
publication.
We also thank the above, as well as L. Spector and H. Bernstein, and
K. Hammerer, G. Vidal, and J. I. Cirac for communicating their results
on classical communications with bidirectional channels.

We are indebted to many colleagues for their inputs to our work.
We thank P. Shor for communicating his RSP results which are crucial 
to our results.
We thank A. Childs and H.-K. Lo for their critical reading of the
manuscript and for many constructive suggestions, part of which
motivated a more precise version of Theorem 1 and the problem on $d_1
\times d_2$ systems.
The finiteness of the Hamiltonian capacities was questioned by
G. Vidal, who also provided the proof for the finiteness of
entanglement capacity.
We thank M. Nielsen for his upper bound on the entanglement capacity
in terms of the Schmidt number.
We thank I. Devetak for important input in proving bound 2.
We thank D. DiVincenzo, J. Dodd, A. Kitaev, B. Terhal, and other
members of the IQI at Caltech for additional helpful discussions.

Since this paper was first posted, other related results have been 
posted~\cite{Berry02a,Childs02e1,Berry02b,Dawson02}.

This work is supported in part by the NSA under the US Army Research
Office (ARO), grant numbers DAAG55-98-C-0041 and DAAD19-01-1-06.

\appendix

\section{Linear bound in communication cost for distillation} 
\label{sec:ccdist}

In this appendix, we obtain a bound on the communication cost in
distillation using \cite{Horodecki01sd} which derives the enhancement
factor of the capacity of a noiseless quantum channel assisted by 
noisy entanglement, i.e. unlimited supply of the mixed state $\rho$.

Suppose given $\rho^{\ot q n}$, $cn$ forward classical bits (in either
direction) is sufficient to distill $n$ ebits ($q \geq 1/D(\rho)$).
Here, we do not require maximum yield of entanglement, so that the
classical communication cost is upper bounded by that required in 
the more difficult job of distillation.

Then, the following is a noisy superdense coding strategy for Alice and
Bob -- first distill and then perform noiseless superdense coding:
\bea
	cn~\mbox{cbits} + \rho^{\ot q n} & \ra & n~\mbox{ebits}
\non
\\
	n~\mbox{ebits} + n~\mbox{qubits} & \ra & 2n~\mbox{cbits}
\non
\eea
Together, the enhancement factor is equal to
${2 n - c n \over n} = 2 - c$, which cannot exceed the optimal value 
\cite{Horodecki01sd} 
\[
	C_{sd} = 1 + \sup_n \sum_{\Lambda_A} 
	\frac{ n S({\rm tr}_A(\rho)) - S(\Lambda_A (\rho^{\ot n}))}
		{S(\Lambda_A({\rm tr}_B (\rho^{\ot n}))}
\]
where the supremum is taken over all trace-preserving completely
positive maps $\Lambda_A$ on Alice's half of $\rho^{\ot n}$.
Hence $c \geq 2 - C_{sd} \equiv \Delta$.  
Even though it is not known how to calculate $\Delta$ for an arbitrary
$\rho$ it is unlikely to be zero for all $\rho$.  If $\exists \rho$
for which $\Delta \neq 0$, then distillation would take at least
linear classical communication.



\end{document}